\documentclass[12pt,preprint]{aastex}

\usepackage{natbib}
\shorttitle{Nature of 3.4 um band in Wild 2 cometary particles and IDPs}
\shortauthors{Matrajt et al.}

\begin{document}


\title{The origin of the 3.4 $\mu $m feature in Wild 2 cometary particles and
 in ultracarbonaceous interplanetary dust particles}


\author{G. Matrajt}
\affil{Astronomy Department, University of Washington,
    Seattle, WA 98195}
\email{matrajt@astro.washington.edu}
\author{G. Flynn}
\affil{Department of Physics, SUNY-Plattsburgh, Plattsburgh, NY 12901}
\author{D. Brownlee}
\author{D. Joswiak}
\and

\author{S. Bajt}
\affil{DESY Photon Science Notkestr. 85 22607 Hamburg Germany}




\begin{abstract}
We analyzed 2 ultra-carbonaceous interplanetary dust particles and 2 cometary Wild 2 
particles with infrared spectroscopy. We characterized the carrier of the 3.4 $\mu$m band 
in these samples and compared its profile and the CH$_{2}$/CH$_{3}$ ratios to the 3.4 $\mu$m band 
in the diffuse interstellar medium (DISM), in the insoluble organic matter (IOM) from 3 
primitive meteorites, in asteroid 24 Themis and in the coma of comet 103P/Hartley 2. 
We found that the 3.4 $\mu$m band in both Wild 2 and IDPs is similar, but different from
 all the other astrophysical environments that we compared to. The 3.4 $\mu$m band in IDPs
 and Wild 2 particles is dominated by CH$_{2}$ groups,  the peaks are narrower and stronger 
than in the meteorites, asteroid Themis, and the DISM.   Also, the presence of the carbonyl group 
C=O at $\sim$1700 cm$^{-1}$ (5.8 $\mu$m) in most of the spectra of our samples,  indicates
 that these aliphatic chains have O  bonded to them, which is quite different from 
astronomical spectra of the DISM.   Based on all these observations
 we conclude that the origin of the carrier of the 3.4 $\mu$m band in IDPs and Wild 2 samples
 is not interstellar, instead, we suggest that the origin lies in the outermost parts of 
the solar nebula. 
\end{abstract}


\keywords{Wild 2 cometary particles; interplanetary dust particles; 3.4 $\mu$m feature}


\section{Aim of paper}
In this work we performed a coordinated study of cometary and interplanetary particles. 
We first used an electron microscope to locate the carbonaceous materials in the samples 
and determine their morphological aspects. We then performed \textit{in situ} infrared spectroscopy 
directly on the carbonaceous materials to investigate their 3.4 $\mu$m band and the presence of 
other organic-related  peaks (carbonyl, aromatics, etc). Finally we compared the 3.4 $\mu$m feature
 of our samples to the 3.4 $\mu$m feature of other astrophysical environments.

\section{Introduction}

The NASA Stardust spacecraft returned  particles collected from the coma of comet 81P/ Wild 2
(hereafter Wild 2). 
 Hundreds of cometary particles ranging from  1 $\mu$m to 100 $\mu$m in
 size were collected  by impact into  aerogel with an encounter velocity of 6.1 km/s \citep{brownlee2006}. 
The examination of these samples has provided many unexpected findings, including the 
presence of refractory minerals \citep{simon2008}, the presence of chondrule-like objects
 \citep{nakamura2008} and low abundance of presolar grains \citep{stadermann2008}. 
Indigenous organic materials are also observed in Wild 2 samples \citep{sandford2006, matrajt2008, cody2008, cody2011,
gallien2008, wirick2009, degregorio2010, degregorio2011, clemett2010, nakamura-messenger2011}.
The abundance of organic matter in the Wild 2 samples was lower than 
expected because the most of the submicron material was destroyed upon capture \citep{brownlee2006}. 
 Coordinated analyses of organic material  have shown that the organic material in the Wild 2 particles is very 
diverse in its morphology, isotopic and chemical composition, abundance, spatial distribution
 and complexity 
\citep{matrajt2008, degregorio2010, degregorio2011, nakamura-messenger2011}.
 Infrared spectroscopy 
 \citep{sandford2006, keller2006,
munozcaro2008, bajt2009}  has shown
 that aliphatics are present in most tracks and particles. In some cases, the aliphatic molecules are
 indistinguishable from the organics intrinsic to aerogel 
\citep{munozcaro2008}. But in most cases
 the organics are very different from the compounds found in aerogel, which is
mainly dominated by 
 CH$_{3}$ groups \citep{bajt2009}.

Interplanetary dust particles (IDPs) are  materials
  collected in the Earth's stratosphere usually considered to be among the most
primitive samples of the solar system \citep{brownlee1976, sandford1987}.
 Most IDPs are very carbon-rich, having in average 10-12 wt${\%}$ C content
 \citep{schramm1989}. The carbonaceous materials in these IDPs are made of organic molecules 
\citep{thomas1993,flynn2003}, including aromatic and aliphatic compounds
 \citep{clemett1993, keller2004}. These carbonaceous
 phases often have H and N isotopic anomalies \citep{messenger2000,
aleon2003, keller2004}
 proving that they are indigenous and suggesting that
they formed through low-temperature chemical reactions \citep{messenger2000,
keller2004, floss2006}  in a presolar cold molecular cloud
 or at the edges of the protoplanetary disk.

In the present study we analyzed with Fourier transform infrared spectroscopy two IDPs and two Wild 2
 samples that have been previously characterized by other analytical \citep{matrajt2008, matrajt2012}. 
These past studies revealed that all of these samples have carbonaceous materials 
with $^{15}$N and D excesses
 and it was suggested that this is primitive organic matter that has changed little or not at all since
 the formation of the Solar System \citep{flynn2003,keller2004,matrajt2008, matrajt2012, matrajt2013}.
    However, owing to their small size, the nature of these phases has been poorly constrained. 
In this work we characterized these organic materials with FTIR to determine 1) the nature of the 
organics; 2) the characteristics of the 3.4 $\mu$m band and 3) the origin of these carbonaceous materials
 (solar $\textit{vs}$ interstellar).

\section{Samples}

In this study we worked with two IDPs, that we nicknamed Chocha and GS
 and two fragments from two Stardust tracks that we nicknamed Febo and Ada.

\subsection{GS}
Particle GS (curatorial name L2055-R-1,2,3,4,5 cluster ${\#}$7) is a “Grigg-Skjellerup timed-collection”  
IDP. Calculations \citep{messenger2002} predicted that 1 to 
50${\%}$ of the total flux of IDPs $>$40 $\mu$m in diameter 
collected after Earth passed through comet 26P/Grigg-Skjellerup's 
dust stream in April 2003 would originate from this comet. A dedicated
  collection of this dust stream was organized by NASA known as 
“Grigg-Skjellerup collection”.  Our sample is an ultra-carbonaceous 
particle made of $>$ 90 ${\%}$ carbon, anhydrous minerals (mainly olivines 
and diopside) and Fe-Mg carbonates. Previous studies of the carbonaceous
 materials of this particle showed that it is composed of several carbonaceous
 textures which have N isotopic anomalies \citep{matrajt2012}.

\subsection{CHOCHA}
Particle Chocha is an IDP  from collector flag W7154. 
It is an anhydrous ultra-carbonaceous particle made of $>$ 95${\%}$ 
carbon. It also contains anhydrous minerals, mainly olivine, 
pyroxene (diopside) and Fe-Ni sulfides (pyrrhotite and pentlandite). 
Previous studies of the cabonaceous materials of this particle showed
 that it is composed of several carbonaceous textures which all have N
 isotopic anomalies \citep{matrajt2012}.

\subsection{FEBO}
Particle Febo is fragment ${\#}$ 2 from Stardust track ${\#}$ 57.
 The particle
 is made mainly of pyrrhotite and fine-grained material and
 also contains small silicates. Previous studies of the carbonaceous
 materials, found in the periphery of the pyrrhotite and between
 the small fine grains, showed that they have several textures 
and N and H isotopic anomalies \citep{matrajt2008}.

\subsection{ADA}
Particle Ada is fragment ${\#}$ 2 from Stardust track ${\#}$ 26. 
The particle is made mainly of tridymite and fayalite. 
Previous studies of the carbonaceous materials, found 
in the periphery of the particle, showed that they have
 several textures and N and H isotopic anomalies (Matrajt et al 2008).

\section{Methoods}

\subsection{Sample preparation}

The Wild 2 particles were received from NASA inside aerogel
chips, also known as keystones. The entire aerogel chips and the IDPs
were embedded in acrylic,   then cut with a diamond knife to
a thickness of less than 50 nm to make it transparent to the electron beam.  
Acrylic was then dissolved out from the cut sections with chloroform vapors,
following the methodology 
developed by \citet{matrajt2006}.

\subsection{Transmission Electron Microscopy (TEM)}
All microtome slices were studied with a 200 keV Tecnai field-emission electron microscope
in transmission mode. 
We used a CCD Orius camera  to study  the morphologies and textures of the carbonaceous 
materials. We also used a Gatan Imaging Filter (GIF) detector  to acquire carbon maps.

\subsection{Fourier Transform Infra Red (FTIR) spectroscopy} \label{bozomath}
Fourier transformed Infrared (FTIR) spectroscopy is a technique often used for
 the $\textit{in situ}$ identification of organic functional groups. The mid infrared 
spectral region, from 650 to 4000 cm$^{-1}$,
 shows unique absorption features
 characteristic of organic materials. 
We used the infrared microscope located on beamline U2B of the National Synchrotron
 Light source at Brookhaven National Laboratory to study samples Febo, GS and Chocha.
  Spectra 
were obtained over a range of 4000 to 650 cm$^{-1}$ and with an energy resolution of 4 cm$^{-1}$
 and a spatial resolution of 3-5 $\mu$m, using a Thermo-Nicolet Continuum FTIR bench 
(KBr beamsplitter) in transmission mode, and a MCT-A detector.
Sample Ada was analyzed at the Advanced Light Source (ALS) 
at Lawrence Berkeley National Laboratory, using a Thermo-Nicolet Magna 760 FTIR bench
 (KBr beamsplitter) and a SpectraTech Nic-Plan IR microscope in reflectance mode, and
 an MCT-A detector. The preliminary IR data of Ada was published in a conference 
abstract \citep{wopenka2008}.

\section{Results}

Table \ref{table1} shows the peak assignments for all the peak positions 
found in all four samples and it also shows which assignments were 
found for each of the samples studied. Table \ref{table2}  shows the CH$_{2}$/CH$_{3}$ 
ratios. Ratios were calculated using the optical depth of the CH$_{3}$ 
peak at $\sim$ 2956 cm$^{-1}$ and the optical depth of the CH$_{2}$ peak at $\sim$ 2926
 cm$^{-1}$ \citep{sandford1991}. We made a baseline correction by fitting
 the baseline with a straight line across the range from 3100 cm$^{-1}$ to
 2800 cm$^{-1}$. Table \ref{table3} shows a comparison of peaks from the 3.4 $\mu$m region 
and the C=O peak  between our samples and other objects (DISM, comet 
103P/Hartley, Murchison IOM, Orgueil IOM, Tagish Lake, asteroid 24 Themis).

The IR peak assignments and interpretations were done based on former IR studies of IDPs \citep{keller2004, matrajt2005,
munozcaro2006} and the Tagish Lake meteorite \citep{matrajt2004} and are as follows: 
3255 cm$^{-1}$ is the OH stretch in water, carboxylic acids
 or alcohols. 
2990  cm$^{-1}$ is a =C-H stretching
2950-2956 cm$^{-1}$ is the CH$_{3}$ asymmetric stretching in aliphatic hydrocarbons. 2918-2920 cm$^{-1}$ 
is the CH$_{2}$ asymmetric stretching in hydrocarbons. 2896 and 2860-2870 cm$^{-1}$ are the CH$_{3}$ symmetric 
stretchings in hydrocarbons. 2845-2855 cm$^{-1}$ is the CH$_{2}$ symmetric stretching in aliphatic hydrocarbons.
2160 cm$^{-1}$ is a
 C=C stretching vibration in alkenes.
1740 cm$^{-1}$ is the carbonyl
 (C=O) in esters.  1730, 1714-1717 and 1700 cm$^{-1}$ are C=O stretching in ketone and carboxylic acids.
1685 cm$^{-1}$ is the H-O-H stretching  in water.
1480  
and 1447 cm$^{-1}$ are CH$_{3}$ and CH$_{2}$ bending vibrations, respectively. 1435 cm$^{-1}$ is C=C 
stretching in aromatics. 1386 cm$^{-1}$ is the CH$_{3}$ symmetric bending. 1270 and 1240 cm$^{-1}$ 
are C-O-C vibrations in esters. 1190 and 1147 cm$^{-1}$ are unknown. 1065 cm$^{-1}$ is a C-OH 
vibration in secondary cyclic alcohol. 987, 970 and 910 cm$^{-1}$ are CH=CH bending vibrations.
Absorption at 1650-1654 cm$^{-1}$ is the C=C stretching in aromatics. 1448 cm$^{-1}$
 is the CH$_{2}$ bending in aliphatics or the CO$^{3-}$ in carbonates.
 1418 cm$^{-1}$ is the C=C stretching in aromatics.
 1350 cm$^{-1}$ is 
the CH$_{3}$ bending in aliphatic hydrocarbons. 1220 cm$^{-1}$ is the CH$_{2}$ wagging mode.
 1160 cm$^{-1}$ is CH$_{2}$ twisting mode.
1216, 1136 and 1106 cm$^{-1}$ are
 Si-O stretching in silicates.
 1070, 1060 and 952 cm$^{-1}$ are the Si-O stretching in pyroxenes.
930 cm$^{-1}$ is the Si-O stretching in silicates and 
887 cm$^{-1}$ is Si-O stretching in olivine.

\subsection{IDP GS}
\subsubsection{TEM}
\begin{figure}
\plotone{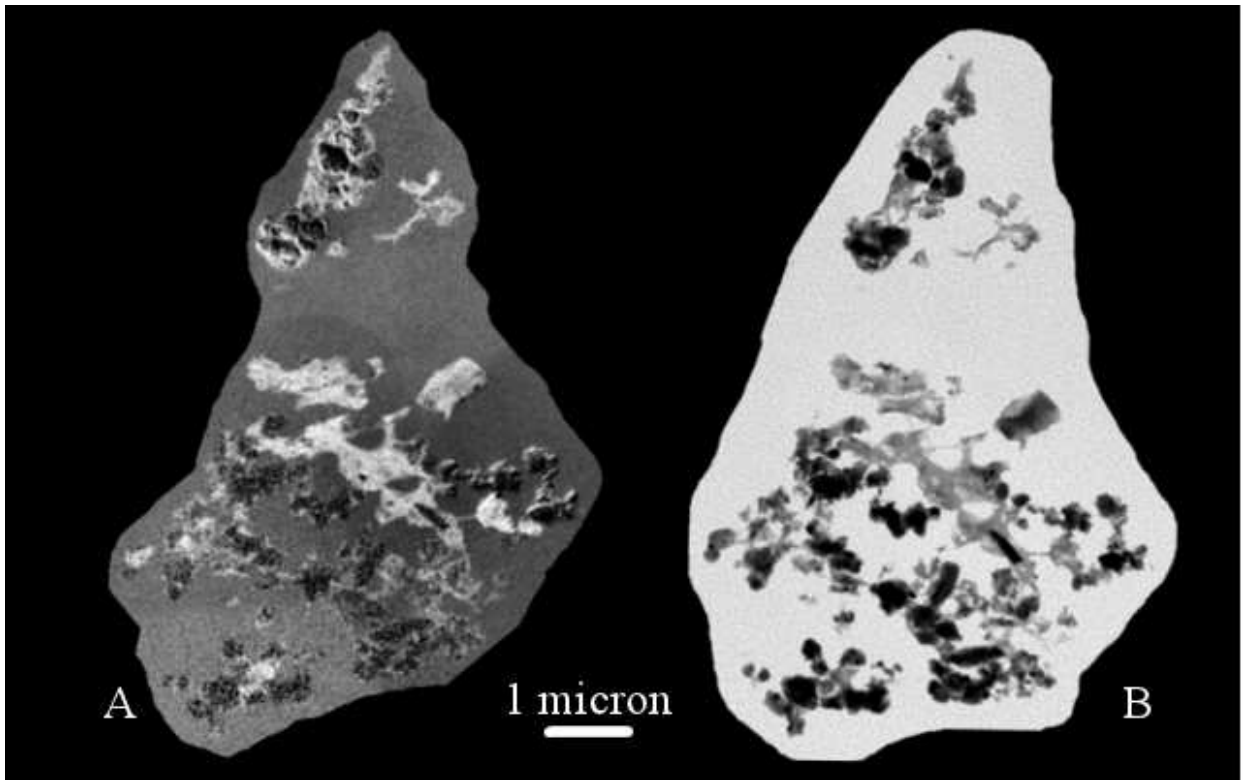}
\caption{Transmission electron micrograph of IDP GS. A) Carbon map of the particle. The bright areas 
are C-rich. B) Bright Field image of the particle.\label{GS-TEM}}
\end{figure}

\begin{figure}
\plotone{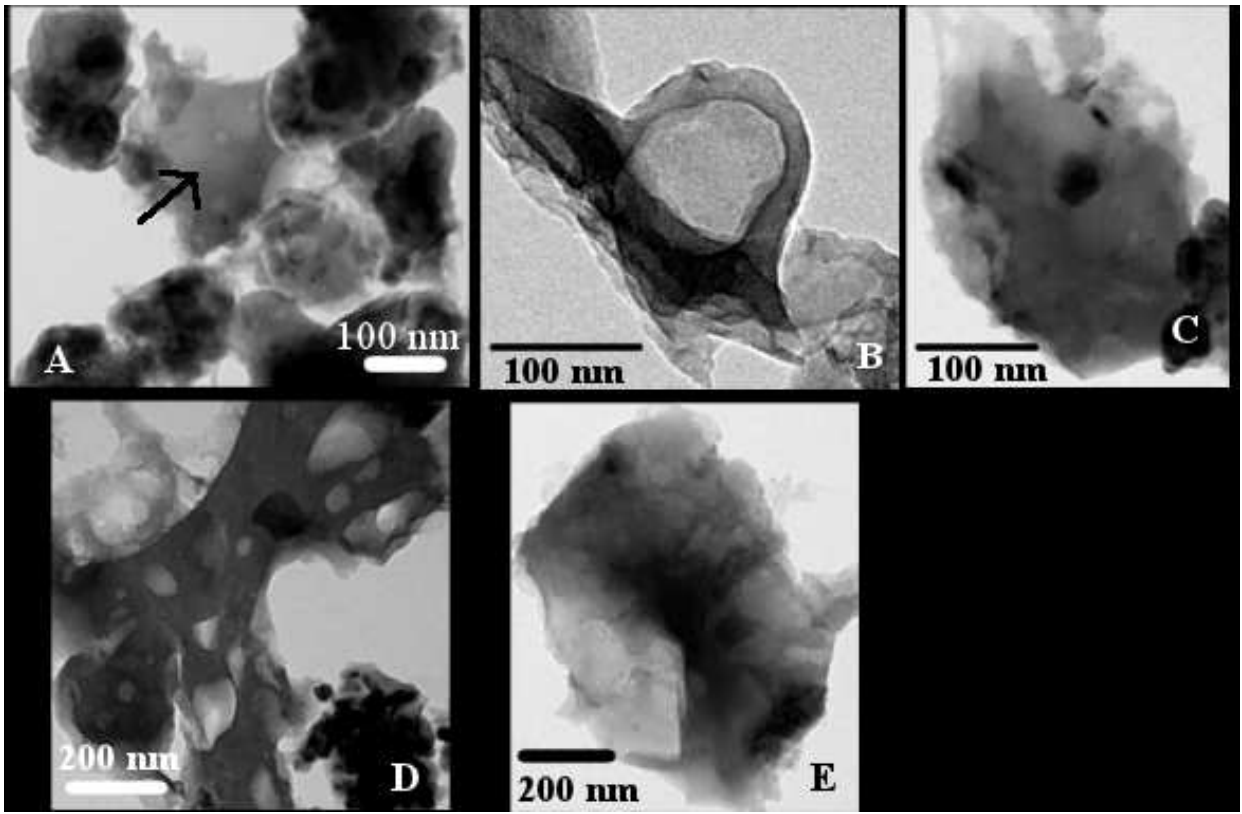}
\caption{Micrographs of IDP GS showing the different textures for the carbonaceous materials.
 A) Vesicular. The arrow points to the small vesicules. B) Globular. C) Dirty.
D) Spongy. E) Smooth. \label{GS-composite}}
\end{figure}

Energy filtered transmission electron microscopy (EFTEM) carbon maps revealed that 
most of the microtomed area of the particle is made of carbonaceous material (Figure \ref{GS-TEM}). 
The bright field (BF) images of the different carbonaceous areas revealed several types
 of morphologies: spongy, globular, smooth, dirty and vesicular. These morphologies, 
previously described in adjacent sections of this same IDP and other IDPs \citep{matrajt2012}
 can be described as follows (Figure \ref{GS-composite}): vesicular morphology  is characterized
 by having small vesicles or voids found in a C-rich smooth material. Usually the voids
 are smaller than the section thickness ($\sim$50-70 nm). Globular morphology is characterized
 by round-shaped structures that may be hollow or filled. Dirty morphology is
 characterized by a carbonaceous material that has mineral grains (typically sulfides)
 embedded in it. Spongy morphology is characterized by a lace mesh-like material.
 Smooth morphology is characterized by a shapeless
 and textureless material.

\subsubsection{IR}
\begin{figure}
\plotone{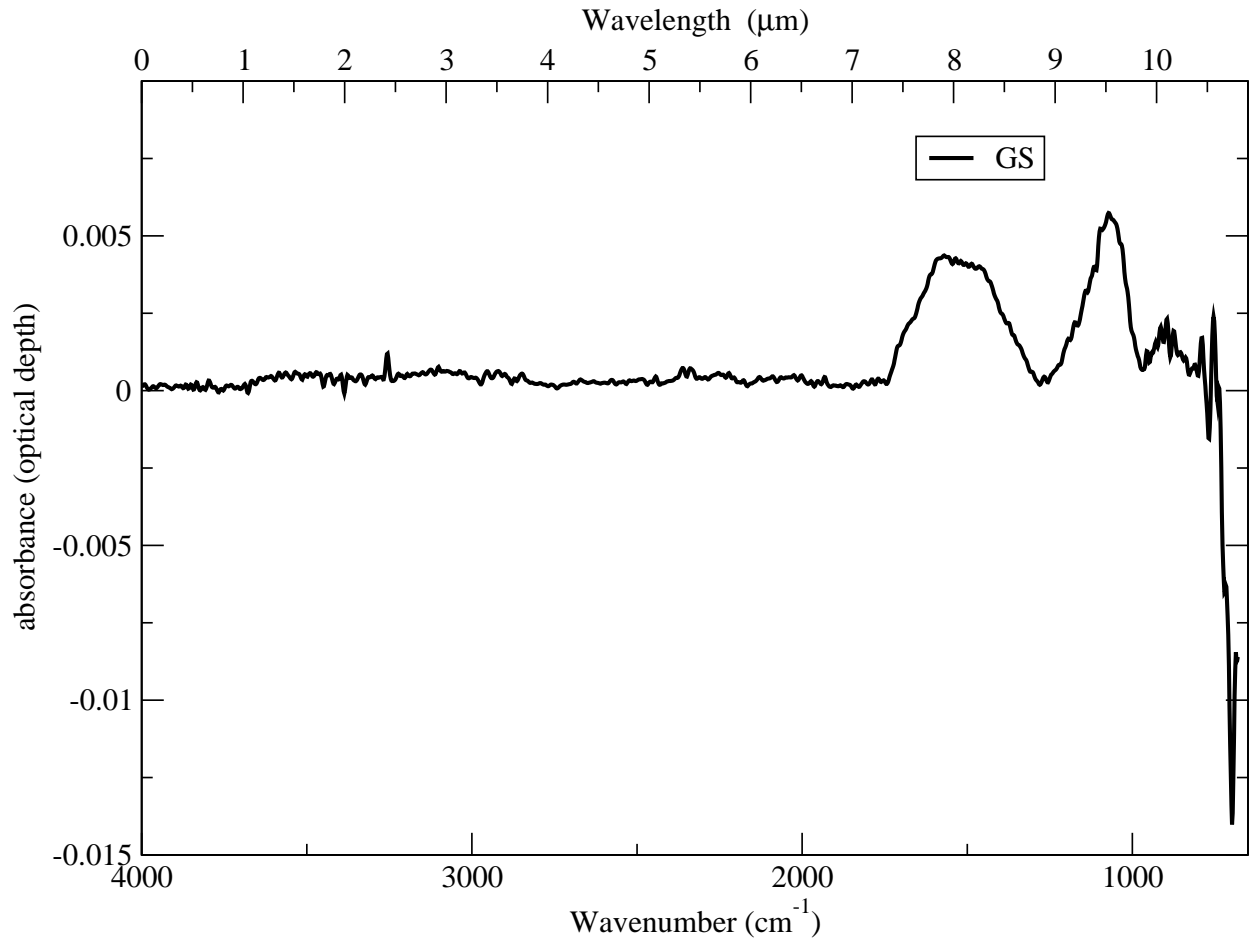}
\caption{IR spectrum of particle GS. \label{GS-absorbance}}
\end{figure}

\begin{figure}
\plotone{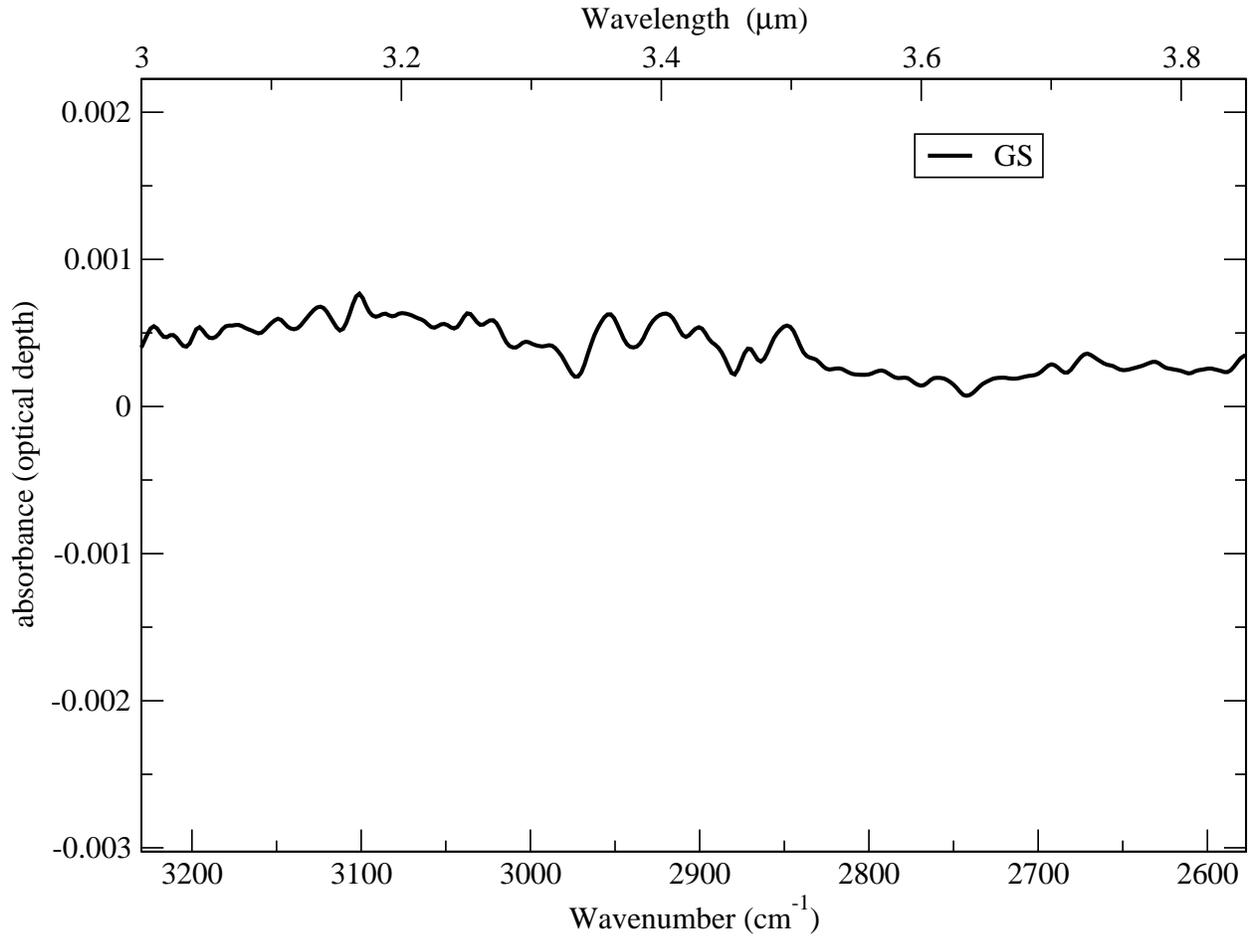}
\caption{IR spectrum of particle GS zoomed in the region $\sim$ 3000 cm$^{-1}$. \label{GS-absorbance-zoom}}
\end{figure}

Figure \ref{GS-absorbance} shows a FTIR spectrum of the entire particle. Peaks are observed at 3255, 2951, 2920, 
2896, 2870, 2845 and 1070 cm$^{-1}$. Also, a broad band from 1545 to 1455 cm$^{-1}$ is observed. From the peak
 assignments we deduced that this particle contains water bonded to its structure, probably to carbonates. It also
contains organics that correspond to aliphatic hydrocarbon chains containing symmetric and assymetric stretchings
and silicates. The broad band from 1545 and 1455 cm$^{-1}$
 corresponds to carbonates. Figure \ref{GS-absorbance-zoom} shows the aliphatic stretching peak area (3000-2800 cm$^{-1}$) 
zoomed. The CH$_{2}$/CH$_{3}$ band depth ratio found was 1.0 (Table \ref{table2}).

\subsection{IDP Chocha}
\subsubsection{TEM}
\begin{figure}
\plotone{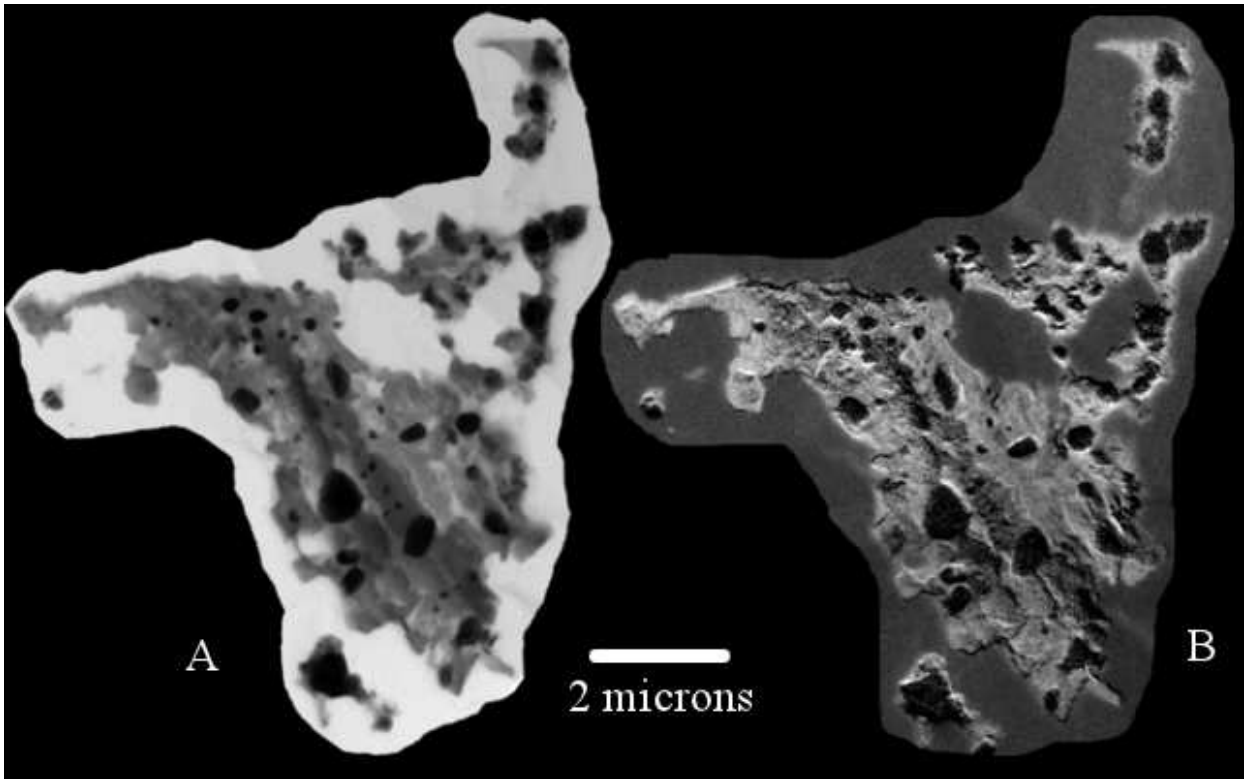}
\caption{Micrograph of IDP Chocha. A) Bright Field image of the particle.  B) Carbon map of particle.
\label{Chocha-TEM}}
\end{figure}

\begin{figure}
\plotone{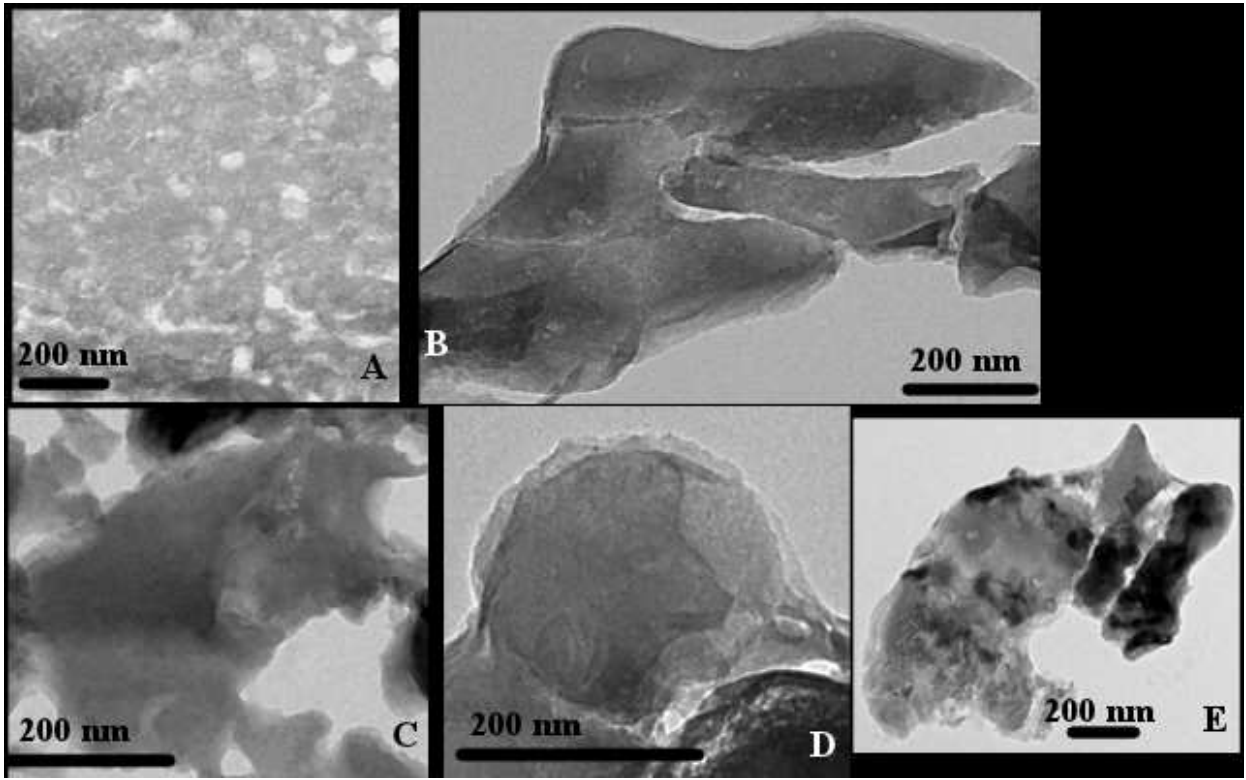}
\caption{Micrographs of IDP Chocha showing the different textures of the carbonaceous
materials. A) Spongy. B) Vesicular. C) Smooth. D) Globular (note this is a filled
globule). E) Dirty.\label{Chocha-composite}}
\end{figure}

EFTEM carbon maps revealed that $>$ 95$\%$
of the microtomed area of the particle is 
made of carbonaceous material (Figure \ref{Chocha-TEM}). The bright field (BF) 
images of the different carbonaceous areas revealed several types of 
morphologies (Figure \ref{Chocha-composite}): spongy , vesicular, smooth,
 globular, and dirty. These
 morphologies were previously described in adjacent sections of this same IDP
 and other IDPs \citep{matrajt2012}.

\subsubsection{IR}
\begin{figure}
\plotone{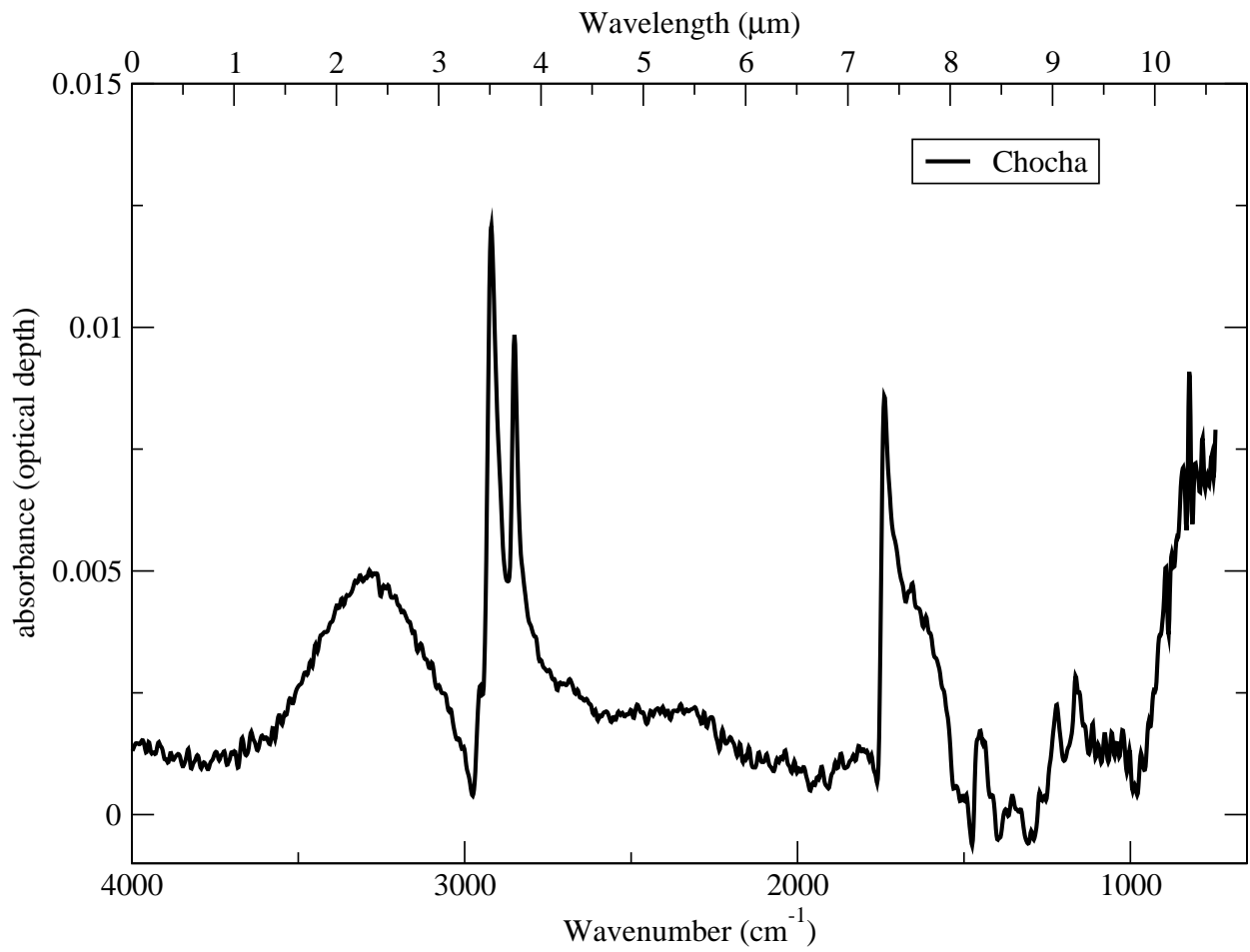}
\caption{IR spectrum of particle Chocha. \label{Chocha-absorbance}}
\end{figure}

Figure \ref{Chocha-absorbance} shows  a FTIR spectrum of the entire particle. 
Peaks are observed at 2956, 2920, 2847, 1740, 1654, 1448, 1350, 
1220 and 1160 cm$^{-1}$. There is also a broad band centered at 3270 cm$^{-1}$. 
From the peak assignments (Table \ref{table1}) we deduced that this particle contains
aliphatic hydrocarbon chains containing symmetric and assymetric stretchings. The organic material also contains carbonyl,
probably in the form of esters, and either olefinic or aromatic C=C molecules. The particle also has
silicates. The broad band is the OH stretch in water. 
The CH$_{2}$/CH$_{3}$ ratio found was 4.6 (Table \ref{table2}). 

\subsection{Wild 2 Febo}
\subsubsection{TEM}
EFTEM carbon maps of the microtomed section revealed several small areas
 that are carbon-rich (Figure \ref{Febo-TEM}). The BF images of the different 
carbonaceous areas reveal several types of morphologies (Figure \ref{Febo-composite}):  dirty , 
vesicular  and smooth. The dirty and vesicular 
morphologies were previously described in adjacent sections of this same 
particle \citep{matrajt2008} and are identical to morphologies identified in
 IDPs \citep{matrajt2012}.

\subsubsection{IR}
\begin{figure}
\plotone{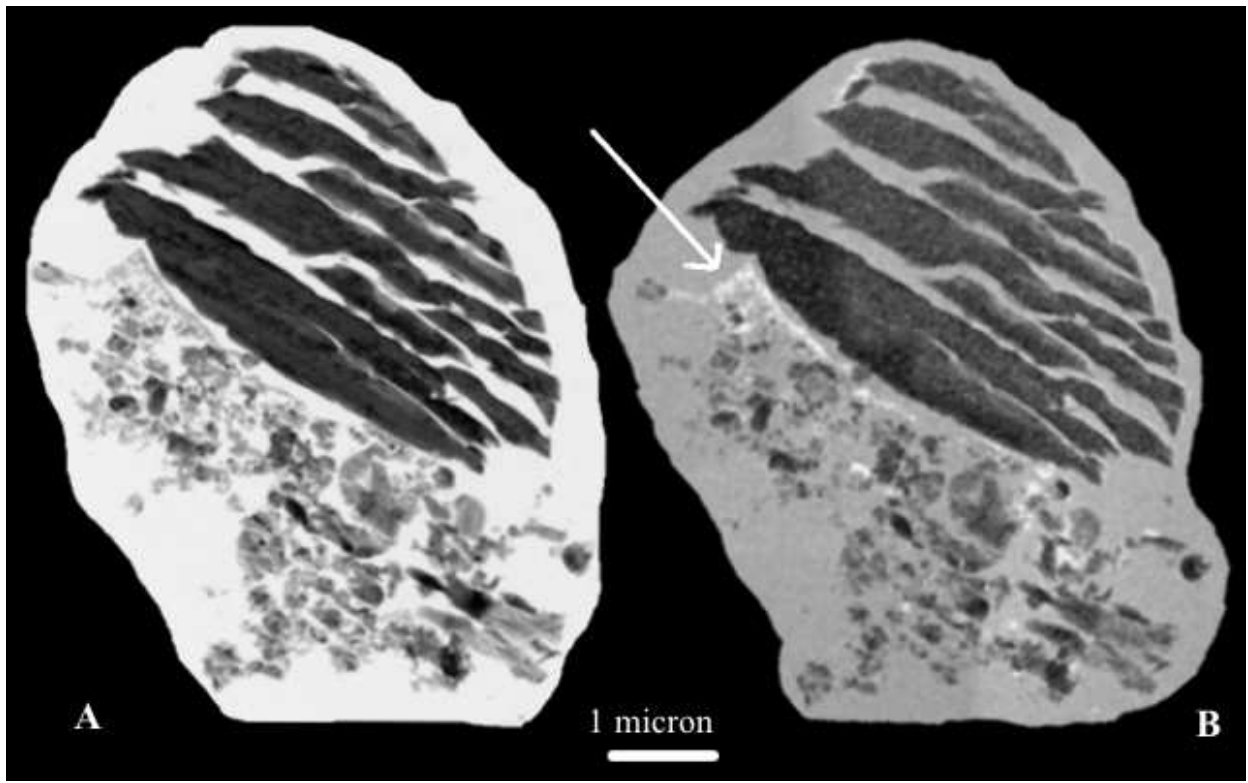}
\caption{Micrograph of Wild 2 particle Febo.  A)
Bright Field image of the particle. B) Carbon map of particle. The arrow
points toward the fine-grained material. \label{Febo-TEM}}
\end{figure}

\begin{figure}
\plotone{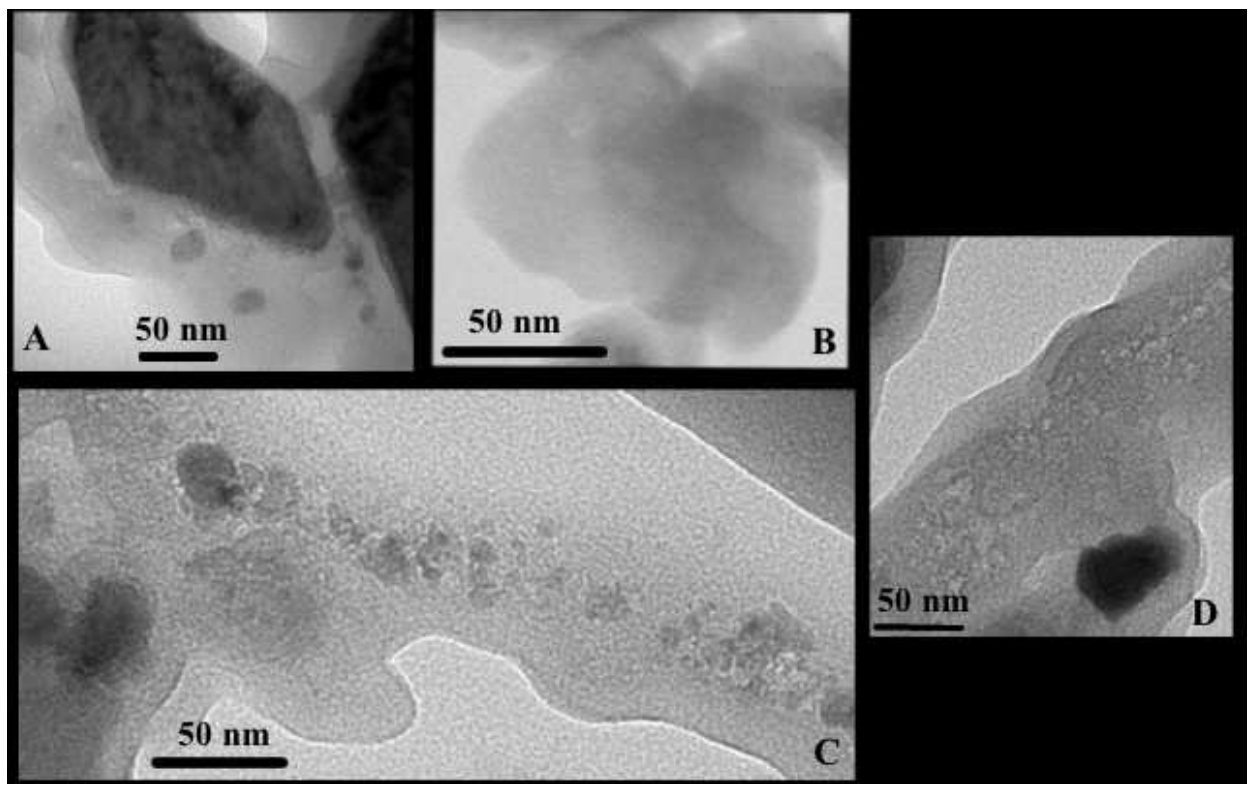}
\caption{Micrographs of particle Febo showing the different textures of the carbonaceous
materials. A) Dirty. B) Smooth. C) Dirty. D) Vesicular.\label{Febo-composite}}
\end{figure}

\begin{figure}
\plottwo{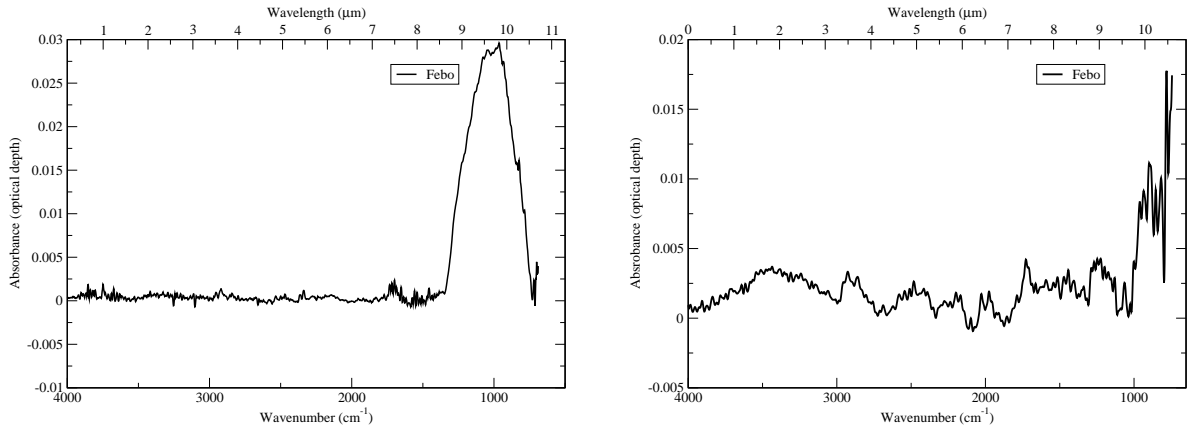}{Febo-area2-absorbance.eps}
\caption{IR spectrum of particle Febo. Left: the sulfide area.
Right: the fine-grained area. \label{Febo-absorbance}}
\end{figure}

Figure \ref{Febo-absorbance} shows two FTIR spectra from two areas of the particle. 
The left one was acquired primarily on top of the sulfide area (black 
area of the particle in figure \ref{Febo-TEM}). The right spectrum was primarily
 acquired from the fine-grained area of the particle (arrow in Figure \ref{Febo-TEM}), which
 is the area where all the C-rich materials were observed. Peaks in the left
 spectrum are observed at 2954, 2920, 2855,  1730, 1717, 1700, 1685 and 1650 cm$^{-1}$. 
There is also a broad band between 1070 and 957 cm$^{-1}$ centered at 1010 cm$^{-1}$. 
The peak assignments indicate that this portion of the particle contains
hydrocarbons with aliphatic chains, carbonyl in ketones and carboxylic acids and water
bonded to the structure of the organic molecules. There is also some evidence of
aromatic or olefinic C=C molecular bonds. 
The broad band is the Si-O stretch in silicates. 

The peaks in the right
 spectrum are observed at 2950, 2920, 2860, 1730, 1060, 952, 930 and 887 cm$^{-1}$. 
The peak assignments indicate that this side of the particle has aliphatic hydrocarbon chains and
carbonyl in ketones and carboxylic acids. There is also evidence of olivines and pyroxenes.
 The CH$_{2}$/CH$_{3}$ band depth ratio was 1.96 (Table \ref{table2}). 

\subsection{Wild 2 Ada}
\subsubsection{TEM}
\begin{figure}
\plotone{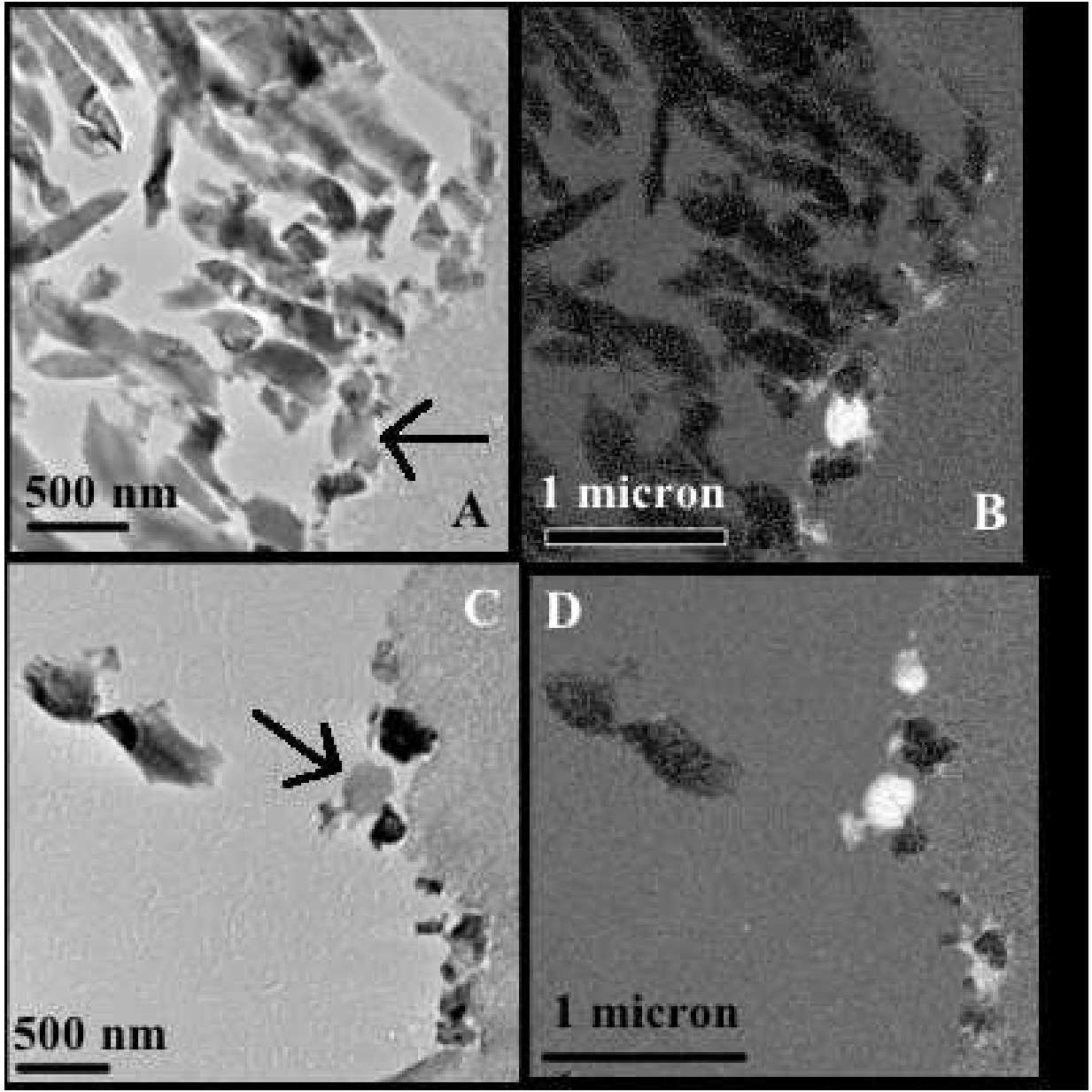}
\caption{Micrographs of particle Ada. A) Bright field of an area of the particle. The arrow points
to the C-rich globule. B) Carbon map of the area shown in A). The bright areas are C-rich.
C) Bright field image of another C-rich area in this particle. The arrow points to a C-rich globule.
D) Carbon map of the area shown in C). \label{Ada-TEM}}
\end{figure}

EFTEM carbon maps of microtomed sections revealed several small areas that are 
carbon-rich (Figure \ref{Ada-TEM}). The BF images of the different carbonaceous areas reveal  
two types of morphologies:  globular  and smooth. 
These morphologies were previously described in adjacent sections of this same 
particle \citep{matrajt2008} and are identical to morphologies identified in IDPs
 \citep{matrajt2012}.

\subsubsection{IR}
\begin{figure}
\plottwo{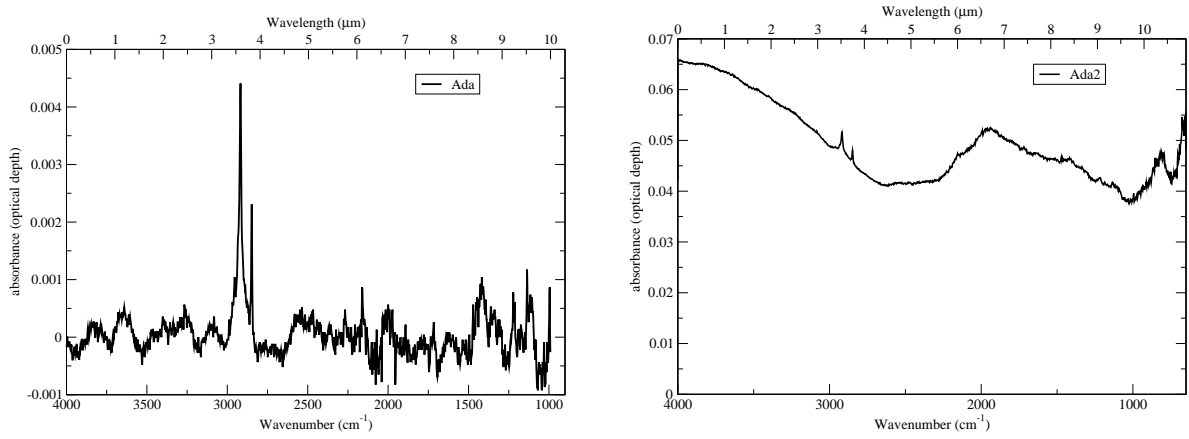}{Ada2-absorbance.eps}
\caption{IR spectra of two areas in particle Ada. \label{Ada-absorbance}}
\end{figure}

Figure \ref{Ada-absorbance} shows two FTIR spectra of two different microtomed sections. Peaks are observed
 at 2954, 2918, 2847, 2160, 1714, 1418, 1216, 1136 and 1106 cm$^{-1}$. Peak assignments
 (Table \ref{table1}) indicate that this particle contains chains of aliphatic hydrocarbons,
some of which have C=C groups attached to them. There is also evidence of carbonyl in ketone
and carboxylic acids and either olefinic or aromatic C=C bonds. Olivines are the main
silicate present in this sample.
The CH$_{2}$/CH$_{3}$ band depth ratio was 4.3 (Table \ref{table2}). 

\subsection{Acrylic}
\begin{figure}
\plotone{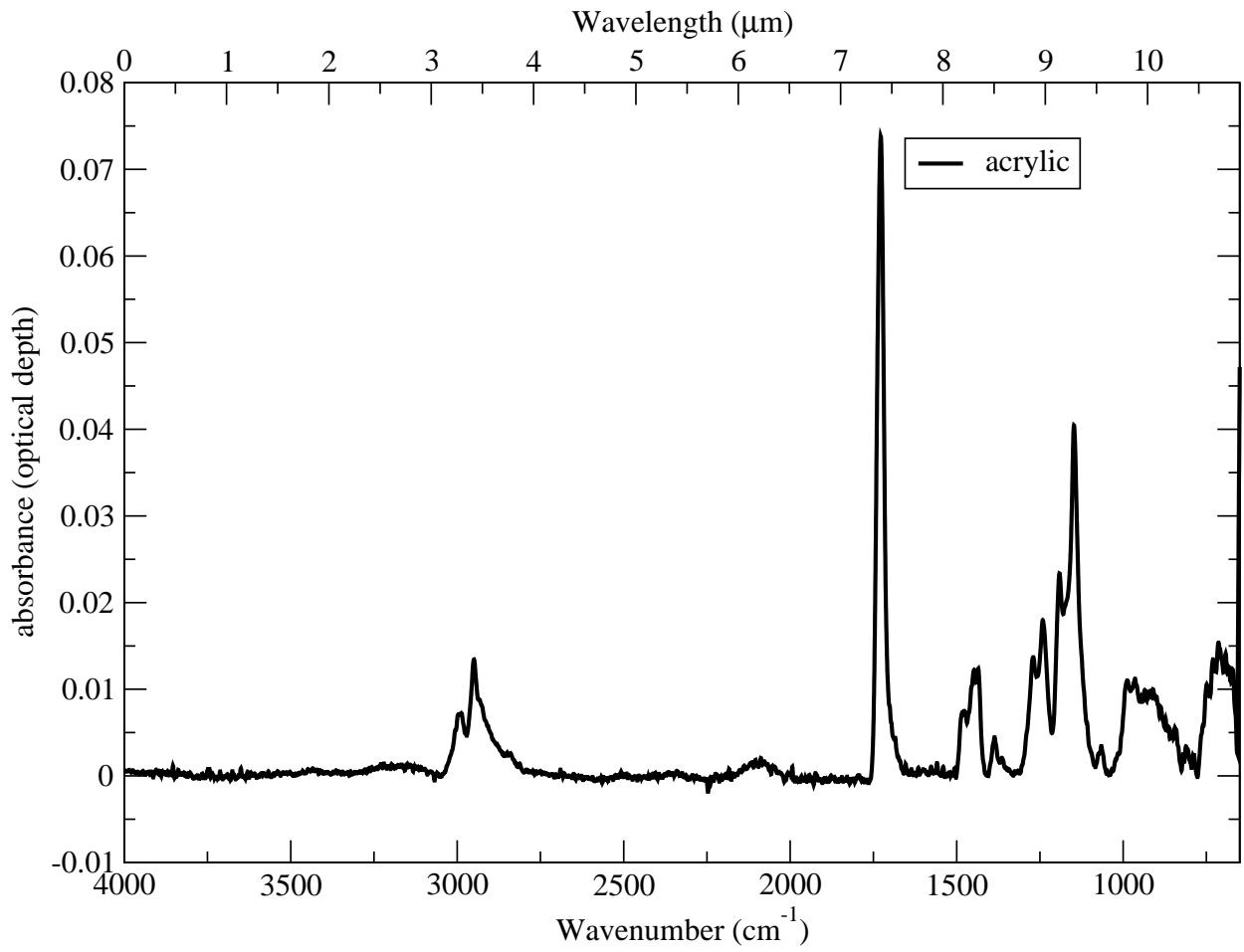}
\caption{IR spectrum of the acrylic embedding medium. \label{Acrylic-absorbance}}
\end{figure}

\begin{figure}
\plottwo{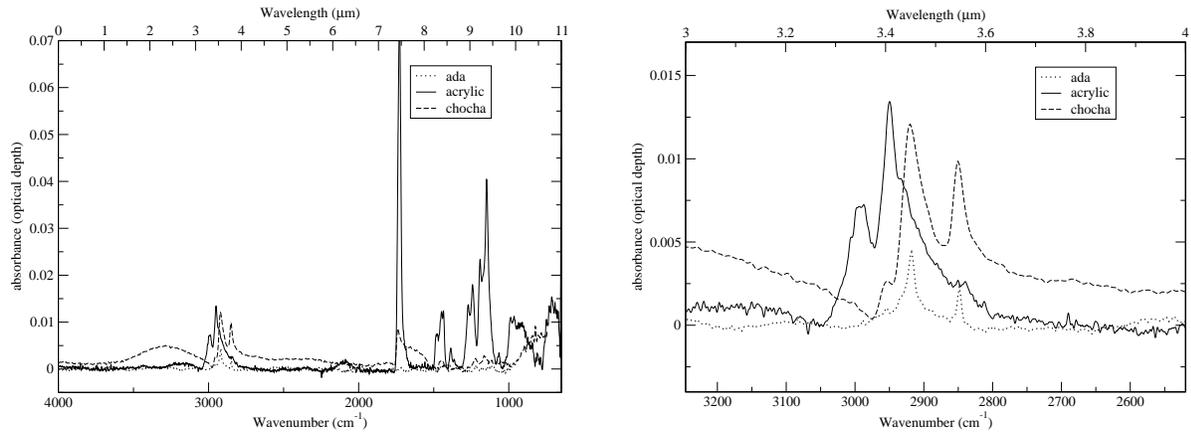}{comparison-zoom.eps}
\caption{Comparison of particles Ada and Chocha to acrylic. Right panel is
a zoom in the 3000 cm$^{-1}$ region of the spectrum shown in the left panel. \label{comparison-absorbance}}
\end{figure}

Although the acrylic embedding medium we used for our samples was removed from 
sections using chloroform vapors, we  measured a piece of acrylic with FTIR under
 the same experimental conditions used for our samples to have a reference spectrum
 and ensure that the interpretations of the organics in the samples are not biased 
by the organics found in acrylic. Figure \ref{Acrylic-absorbance} shows a FTIR spectrum of this acrylic.
 The peaks observed are at 2990, 2949, 1727, 1480, 1447, 1435, 1386, 1270, 1240, 1190, 
1147, 1065, 987,  970 and 910 cm$^{-1}$. 
Peak assignments (Table \ref{table1}) indicate that acrylic is composed of
aliphatic hydrocarbons, ketones and carboxylic acids, aromatics and esters.
There are also secondary cyclic alcohols.
Figure \ref{comparison-absorbance} shows a comparison of acrylic with IDP Chocha and Wild 2 particle Ada. 
The spectrum of acrylic is very different from the other two spectra. First, the peaks in
 the 3000 cm$^{-1}$ region are shifted in the acrylic toward higher values (2995 and 2950 cm$^{-1}$), 
while both in Chocha and Ada these peaks are in similar positions and shifted toward lower 
values (2917 and 2848 cm$^{-1}$). Acrylic has a C=C-H stretching that is absent in Ada and Chocha.
 Acrylic is dominated by CH$_{3}$ while Ada and Chocha are dominated by CH$_{2}$ groups (Table \ref{table1}). 
Second, the relative heights of these peaks are very different in the acrylic spectrum. 
Third, Ada, Chocha and acrylic have a peak around 1700 cm$^{-1}$, but in the 
acrylic this peak is very narrow and strong comparing to the one found in our samples and its position
 is slightly shifted to lower values comparing to our samples. Additionally, sample GS lacks
a peak at this position (Figure \ref{GS-absorbance}), clearly indicating that the presence of this peak
in our samples is not related to the acrylic embedding medium. Forth, all the peaks below 1500
 cm$^{-1}$ in the acrylic spectrum are narrower and stronger (more intense) than in Chocha (and
 they are absent in Ada). Fifth, some peaks in the acrylic spectrum (in the 1000 cm$^{-1}$ region)
 are absent in our samples. In general, peaks in the acrylic spectrum are better defined (more net)
 and narrower and stronger than in the samples we studied. It is clear that the organic material
 measured in our samples is unambiguously indigenous to the particles and not a contamination
 from the acrylic embedding medium. The CH$_{2}$/CH$_{3}$ band depth ratio could not be calculated because
 acrylic does not have a CH$_{2}$ peak at ~2920 cm$^{-1}$.

\section{Discussion}
The textures and morphologies of the carbonaceous materials we found in our samples are
 identical to the morphologies found in carbonaceous materials from carbonaceous chondrites
 \citep{garvie2004, garvie2006, nakamura-messenger2006}, from interplanetary dust particles
 \citep{matrajt2012} and from other Wild 2 particles \citep{matrajt2008, degregorio2010, matrajt2013}.
 These carbonaceous materials are organic refractory
 molecules, given that they survive atmospheric entry or hypervelocity impact into aerogel
 \citep{matrajt2012, matrajt2013}. Both IDPs and Wild 2 samples suffered from heating 
while being decelerated either in the stratosphere or aerogel. The effects of heating on
 the organics in these type of samples are poorly known. However, past studies have shown
 that pyrolyzed terrestrial kerogens tend to increase their CH$_{3}$/CH$_{2}$ ratios as well as their
 degree of aromatization \citep{ehrenfreund1991}. For example,  under the effects of heating the
 Orgueil meteorite decreased its CH$_{3}$/CH$_{2}$ ratio. It was suggested that
 such a decrease means that the -CH$_{3}$ groups are engaged in thermally labile structures ($\textit{i.e.}$ bounded
 to N and/or O). This would cause a faster decomposition of CH$_{3}$ groups relative to the CH$_{2}$ groups
  and would change the CH$_{3}$/CH$_{2}$ ratios  \citep{ehrenfreund1991}. It could also be due simply to
  the general loss of H from these materials during pyrolysis \citep{jones2012I}.
This is, however, not the case for the samples
 analyzed in our study. Both IDPs and Wild 2 particles contain abundant O and N bonded to their
 organic materials \citep{flynn2003, matrajt2005, sandford2006, matrajt2008, degregorio2010, matrajt2012, matrajt2013}
 indicating that the CH$_{3}$ groups did not 
decompose by outgassing of thermally labile structures during deceleration. CH$_{3}$ groups are simply
 less abundant in IDPs and Wild 2 samples relative to other astrophysical environments. Therefore,
 we believe that the CH$_{2}$/CH$_{3}$ ratios discussed in the following sections reflect the primary composition
 of the organic molecules from the parent bodies of our samples.

In the following paragraphs we will discuss the characteristics of the 3.4 $\mu$m feature of this organic
 refractory material and we will compare this feature to the one observed in the interstellar medium, 
meteoritic material from carbonaceous chondrites (CCs), comet Hartley 2 and asteroid Themis. Because 
infrared spectroscopy is primarily a qualitative analytical technique, these comparisons  will remain 
purely qualitative.

The meteoritic material that has been previously investigated with infrared spectroscopy consists of 
two different components. First, a general carbonaceous component of the meteorite, which consists of 
all the carbonaceous materials present in the sample.  Second, a residue component, which consists of 
carbonaceous materials that are water insoluble and are usually termed $\textquoteleft$$\textquoteleft$ insoluble organic matter$\textquotedblright$ (IOM).
 This residue is obtained by a series of acid treatments that involve several chemicals including 
hydrochloric acid (HCl) and hydrofluoric acid (HF). These treatments are designed to concentrate the
 carbonaceous material by dissolving the silicates \citep{reynolds1978, alexander1998}. The general
 carbonaceous component was investigated mainly in the Tagish Lake meteorite \citep{matrajt2004}, whereas
 the acid residue component was investigated in the Orgueil and Murchison meteorites \citep{ehrenfreund1991,
devries1993, flynn2003, flynn2010}. In the discussion that follows, we will refer to the IOM of Murchison and Orgueil
 meteorites. However, for Tagish Lake we will refer to Tagish Lake as the general carbonaceous component.
 This distinction is important given that it was recently demonstrated that exposing organic
 matter to HF and HCl,  alters significantly the organic molecules 
  by increasing the aliphatic CH$_{3}$/CH$_{2}$ ratio and changing the aromatic
contents  \citep{flynn2010}.

\subsection{The 3.4 $\mu$m band}
The C-H stretch region in the IR spectrum is also known as the 3.4 
$\mu$m band. Past observations
 of several lines of sight of the diffuse interstellar medium \citep{sandford1991,
pendleton1994, chiar2000}
 have shown that this 3.4 $\mu$m region shows strong similarities with the 3.4 
$\mu$m band of the insoluble organic matter (IOM) of meteorites Orgueil and Murchison 
\citep{ehrenfreund1991, devries1993}.
 One way to directly compare the spectra of the DISM with meteoritic materials,
 including Wild 2 samples and IDPs, is through the comparison of the CH$_{2}$/CH$_{3}$ band depth ratio. The peaks
 at $\sim$2956 cm$^{-1}$ and $\sim$2926 cm$^{-1}$ correspond to the asymmetric stretching of
 CH$_{3}$ and CH$_{2}$, respectively. 
These peaks are almost always present in the IR spectra of both the DISM and meteoritic materials. By 
constructing a linear continuum  across the range from about 3000 cm$^{-1}$ to 2800 cm$^{-1}$, as it was done by
 \citet{sandford1991}, we can derive the band depths of these two peaks and then
calculate their ratios. This technique has been used extensively in the past for comparison of the DISM
 to different types of organic materials, whether extraterrestrial \citep{sandford1991,
ehrenfreund1991, devries1993, flynn2003, matrajt2004, matrajt2005, keller2006,
munozcaro2008}  or terrestrial analogs \citep{sandford1991, ehrenfreund1991, 
pendleton1994, pendleton2002, duley1998}.  We calculated the ratio for the four samples
 we analyzed and compared them with the published data of other samples, including CCs,  the DISM, 
comet Hartley 2 and asteroid 24 Themis (Table \ref{table2}). Given that the carrier of the 3.4 $\mu$m interstellar 
absorption band appears to be widespread throughout the Galaxy and is present in several others
\citep{pendleton1997, wright1996, imanishi2000, dartois2004, dartois2011, godard2012}, 
we used the interstellar CH$_{2}$/CH$_{3}$ ratio measured in the past by 
\citet{sandford1991}, \citet{pendleton1994} and \citet{pendleton2002} 
 for the comparison. Another tool that we used for the comparison is the presence of O and N
 in the organic materials. The organic material of the diffuse interstellar medium is mostly
 hydrocarbon in nature, having little N or O, with carbon distributed between  aromatic
 and aliphatic molecules and no bands in the 1000-2000 cm$^{-1}$ region
\citep{pendleton2002}, while the organic
 refractory material in CCs contains O and N 
\citep{devries1993, matrajt2004}. 

\subsubsection{The 3.4 $\mu$m feature of Wild 2 particles}

Wild 2 particle Febo has carbonaceous morphologies that slightly differ from Wild 
2 particle Ada. Although both particles have smooth and globular morphologies 
\citep{matrajt2008}, Febo has also dirty and vesicular morphologies which are 
absent in Ada. Both particles have several similarities in their IR spectra. 
For example, both particles have the CH$_{3}$ symmetric stretch ($\sim$2955 cm$^{-1}$), the CH$_{2}$
 stretch (2918 and 2847 cm$^{-1}$) and the C=O stretch ($\sim$1710 cm$^{-1}$). They also have C=C 
aromatic stretching although at different positions (1418 and 1650 cm$^{-1}$). The main 
difference is that the CH$_{3}$ peaks are slightly more pronounced in Febo than in Ada 
and the CH$_{2}$ peaks are narrower and stronger in Ada. This indicates that CH$_{2}$ groups 
are more dominant in Ada than in Febo and so the aliphatic chains are shorter and/or 
more branched in particle Febo. These morphological and spectroscopic differences 
suggest that comet Wild 2 has a heterogeneous composition. The organic carbonaceous
 materials vary at the track scale (few microns). Similar conclusions were found in
 past studies related  to isotopic compositions of Wild 2 particles where it was found
 that Wild 2 particles may contain carbonaceous materials with isotopic anomalies 
adjacent to carbonaceous materials with normal isotopic compositions \citep{matrajt2008,
degregorio2010, degregorio2011, nakamura-messenger2011, matrajt2013}.
The ratio of the band depth of the CH$_{2}$/CH$_{3}$ peaks was calculated for particle Febo (Table \ref{table2}), 
and it is 1.96. This ratio is within the range found for IDPs 
\citep{flynn2003, matrajt2005}, and the ratios measured in other Wild 2 particles
\citep[Table \ref{table2}]{keller2006, sandford2006, munozcaro2008}.
  The CH$_{2}$/CH$_{3}$ ratio of Febo and other Wild
 2 samples is slightly larger than the ratios (1-1.5) measured in organic materials from the
 IOM of  CCs (Table \ref{table2}) and the ratios (0.96-1.25) measured in several lines of sight of the
 diffuse interstellar medium of our galaxy \citep{sandford1991, pendleton1994,
chiar2000}. This indicates that the aliphatic molecules responsible of the 3.4 $\mu$m band are
 longer and/or less branched in Wild 2 than in the IOM of  CCs and the DISM and suggests that
 the origin of the organic carbonaceous materials in Wild 2 is not the same as in  CCs, believed
 to be interstellar processes such as UV photoirradiation of ices \citep{munozcaro2008}. The
 ratio of the band depth of the CH$_{2}$/CH$_{3}$ peaks was also calculated for particle Ada 
(Table \ref{table2})
 and it is 4.3. This ratio is within the range found for IDPs \citep{flynn2003} but considerably
 larger than the ratios measured in other Wild 2 particles \citep{keller2006,
sandford2006, munozcaro2008}. This difference is most likely related to the heterogeneity of the organic
 materials of comet Wild 2. The molecules responsible for the 3.4 $\mu$m band in particle Ada are clearly
 different from the molecules making up the 3.4 $\mu$m band in the IOM of  CCs and DISM as the CH$_{2}$/CH$_{3}$
 ratio is 3-4 times larger, indicating that the chains are longer and less ramified in Ada and, as
 stated above, that their origin is not interstellar. As it was suggested in previous studies 
\citep{sandford2006}, most likely the Wild 2 organics are the result of protosolar nebular processes.

\subsubsection{The 3.4 $\mu$m feature of interplanetary dust particles}

Particles GS and Chocha have similar carbonaceous morphologies and these account for 
the same surface area in both particles ($\sim$90$\%$). 
However, they are quite different between themselves with respect to their 
IR spectra (Figures \ref{GS-absorbance} and \ref{Chocha-absorbance}). 
While GS has small CH peaks, Chocha has stronger and narrower peaks. 
The positions of the peaks have some similarities between the two IDPs. For example, both 
particles have CH$_{2}$ stretching ($\sim$2920, 2847 cm$^{-1}$), both samples have a OH stretching
 ($\sim$3260 cm$^{-1}$)
 indicating the presence of water in their structure. This water might be bonded to carbonates, 
carboxylic acids or alcohols. Carbonates were observed by TEM in particle GS. Particle Chocha does
 not have carbonates, therefore water might be part of the structure of its organics inventory
 ($\textit{i.e.}$ alcohols, esters, etc). There are also several differences in the positions of the peaks.
 First, particle Chocha has a very small asymmetric CH$_{3}$ stretching ($\sim$2955 cm$^{-1}$) and the symmetric
 CH$_{3}$ stretching ($\sim$2865 cm$^{-1}$) is absent. The asymmetric CH$_{2}$ stretching ($\sim$2847 cm$^{-1}$)
 is present in
 both particles but in Chocha this is a much narrower and stronger peak. Additional peaks corresponding
 to CH$_{2}$ bending and symmetric stretching ($\sim$1448, 1220 and 1160 cm$^{-1}$) and to C=O and C=C stretching
 (1740 and 1654 cm$^{-1}$) are absent in particle GS but are present in particle Chocha (Table \ref{table1}).
 The ratio
 of the band depth of the CH$_{2}$/CH$_{3}$ peaks was calculated for both IDPs (Table \ref{table2}).
 The value found for GS
 is 1.01, within the  IDP range (1.0-5.6) found in past studies \citep{flynn2003,
matrajt2005}.
 The value found for Chocha was 4.6, also within the IDP range (Table \ref{table2}). The aliphatic molecules making
 the 3.4 $\mu$m band of Chocha are dominated by CH$_{2}$ groups, and there is also evidence of OH and C=O groups,
 indicating that the molecules are made of long chains with OH and C=O groups branched to them. The profile of the
 IR spectrum of these organics looks quite different from 
 the IOM of CCs and the DISM, suggesting that they did not originate from interstellar processes.

Both IDPs and Wild 2 particles seem to have several similarities in their carbonaceous materials. 
Morphologically, all 4 samples studied here have globular and smooth morphologies. Although all four
 particles have CH$_{2}$ stretching peaks and few CH$_{3}$ stretching peaks,  in general CH$_{2}$ dominates
 over CH$_{3}$.
 All but GS have C=O carbonyl stretching and C=C aromatic peaks (Table \ref{table1}). Only the IDP samples
 show bonded water (broad band at $\sim$3250 cm$^{-1}$). These observations indicate that the organic materials
 in the parent bodies of IDPs are similar to Wild 2, with a propensity to have long, less ramified aliphatic
 chains (CH$_{2}$ dominance), with some carbonyl and aromatic groups attached to them. However, the cometary
 particles have less hydrated phases (absence of hydrated minerals like carbonates and absence of OH
 stretching, Table \ref{table1}) than the IDPs, which might be related to the way they were collected (hypervelocity
 impact in aerogel).

\subsubsection{The meteorite 3.4 $\mu$m feature}

The 3.4 $\mu$m feature has been studied in several carbonaceous chondrites.  Murchison and Orgueil 
were among the first to be investigated \citep{ehrenfreund1991, flynn2003,
flynn2010}. 
The Tagish lake meteorite was also investigated \citep{matrajt2004}. The 3.4 $\mu$m feature in the
 IOM of both Murchison and Orgueil are very similar and are also quite similar to the DISM 
\citep{ehrenfreund1991}. In these objects and astrophysical environments the CH$_{2}$ and CH$_{3}$ 
peaks are comparable in both their positions, shapes and relative depths, whereas the 3.4 $\mu$m
 feature in Tagish lake is dominated by CH$_{2}$ groups \citep{matrajt2004}. Also, in Tagish Lake
 the CH$_{2}$ peaks are narrower than in the IOM of the other two meteorites and the DISM. The position
 of the peaks in these three meteorites are similar to the ones found in our samples (Table \ref{table3}). However,
 the shape of the peaks in Murchison and Orgueil IOM differ from our samples given that in our samples,
 specially in Chocha and Ada, the CH$_{2}$ peaks are much narrower and stronger relative to the CH$_{3}$ peaks.
 The 3.4 $\mu$m band of Tagish Lake has a similar profile than the 3.4 $\mu$m band in our samples, where CH$_{2}$
groups dominate over CH$_{3}$ groups. The ratio CH$_{2}$/CH$_{3}$ of the IOM in Murchison has been measured 
independently by several groups \citep{ehrenfreund1991, flynn2003}. The value measured varies
 between 1.09 and 1.51. This range is probably related to the heterogeneity in the distribution of organic
 matter in Murchison \citep{pizzarello2004}. In general the CH$_{2}$/CH$_{3}$ ratios in Wild 2 and IDPs are higher
 than in the IOM of Murchison and Orgueil (except for particle GS), suggesting that the carrier in these
 two CCs  has more ramifications than the carriers in Wild 2 and IDPs. The 3.4 $\mu$m band in both Orgueil and
 Murchison IOM is very similar. However, it is quite different from the Tagish Lake band. This difference
 might be attributed to the fact that Tagish Lake is an unclassified meteorite \citep{brown2000} with ~5.8 wt${\%}$
 of carbon \citep{grady2002}, two-three times as much as it is found in Murchison and Orgueil ($\sim$2 and 3 wt${\%}$, 
respectively) \citep{oro1971, kerridge1985}. However, we favor the idea from \citet{flynn2010} that the 
acid extraction treatment alters the organic materials and thus this alteration might be the explanation for
 the differences seen and described above in the 3.4 $\mu$m band between Tagish Lake and the IOM of Murchison and
 Orgueil.

\subsubsection{The 3.4 $\mu$m feature in asteroids and comets}

\begin{figure}
\plottwo{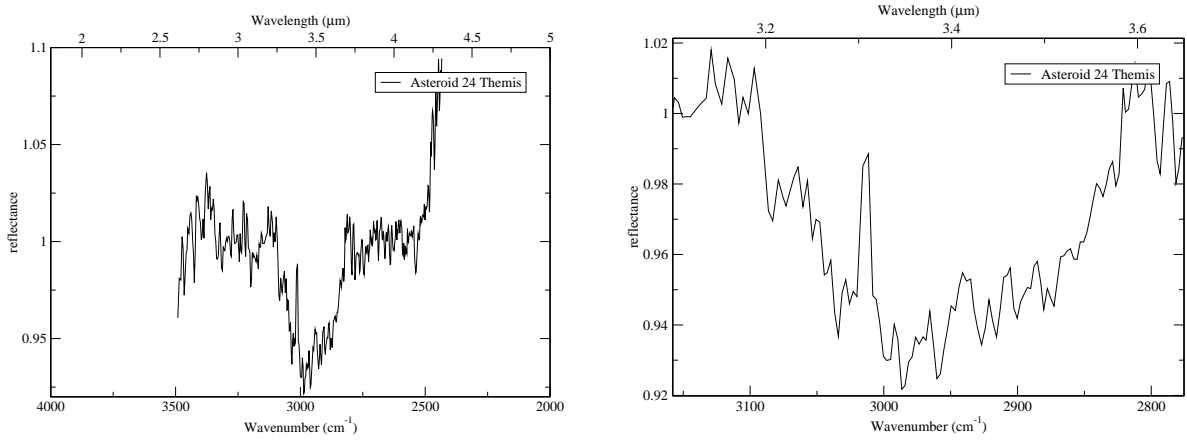}{themis-zoom}
\caption{Left panel: IR spectrum of asteroid 24 Themis ratioed to 
the water-ice model of \citet{rivkin2010}. Right panel: zoom of the spectrum
shown in the left panel. Spectrum reproduced from \citet{rivkin2010} and \citet{campins2010}. \label{themis-IR}}
\end{figure}

Organics have been detected in asteroid 24 Themis, an object with cometary characteristics \citep{hsieh2006}
 that belongs to the same dynamical family as three of the five known Main Belt Comets (MBCs) 
\citep{campins2010, rivkin2010}. The infrared spectrum obtained from these observations
 has been reproduced with permission from the respective authors in Figure \ref{themis-IR}. The general
 shape of the band
 extending from 2500 to 3500 cm$^{-1}$ and centered at  3000 cm$^{-1}$ is quite different from the 3.4 $\mu$m 
band of
the DISM, IOM of Murchison, Tagish Lake and our samples. The authors attributed this band to fine-grained
 water ice as a frost deposited on regolith grains. However, the spectral structure between 2700 and 3000
 cm$^{-1}$ cannot be explained by the presence of ice. Therefore, the authors performed simulations by including
 organics in their models \citep{campins2010, rivkin2010}. The spectrum we reproduced in Figure \ref{themis-IR}
 is in fact the 24 Themis spectrum ratioed to the water-ice model of \citet{rivkin2010}. 
 Although the band is quite saturated,
 many individual peaks can be distinguished (see zoom in Figure \ref{themis-IR}). 
They have positions at 2856, 2876, 2915, 2926, 2946, 2958 and
 2994 cm$^{-1}$. Many of these peaks are also found in the DISM and our samples (Table \ref{table3}) but
 the overall profile
 of this spectrum is quite different from meteoritic materials (including our samples) and DISM. The presence
 of these peaks suggests that aliphatic chains containing CH$_{3}$ and CH$_{2}$ groups are present in
 the 3.4 $\mu$m band
 of all these astrophysical environments, but the depth and general shape of these peaks is quite different
 from the DISM and meteoritic materials indicating that the compounds that make the 3.4 $\mu$m band in 24 Themis
 are different from the ones in meteoritic materials and the DISM. For example, the weakness of the peak at
 2856 cm$^{-1}$ (CH$_{2}$) relative to the peak at 2958 cm$^{-1}$ (CH$_{3}$) in the spectra of 24 Themis 
when compared to similar
 peaks in the spectra of Ada (Figure \ref{Ada-absorbance}) or Febo (Figure \ref{Febo-absorbance}),
 suggests that Wild 2 is richer in -CH$_{2}$ groups than 24 Themis.
 The same comparison between the spectrum of 24 Themis and the IDPs Chocha (Figure \ref{Chocha-absorbance})
 and GS (Figure \ref{GS-absorbance}) also shows that
 IDPs are richer in CH$_{2}$ groups than 24 Themis. Due to over-saturation of peaks in the spectrum of 24 Themis we
 did not calculate the CH$_{2}$/CH$_{3}$ ratio. However, from the general profile of the band in 24 Themis we 
conclude that
 the organics in our samples are very different from the organics that make the 3.4 $\mu$m band in 24 Themis.

\begin{figure}
\plotone{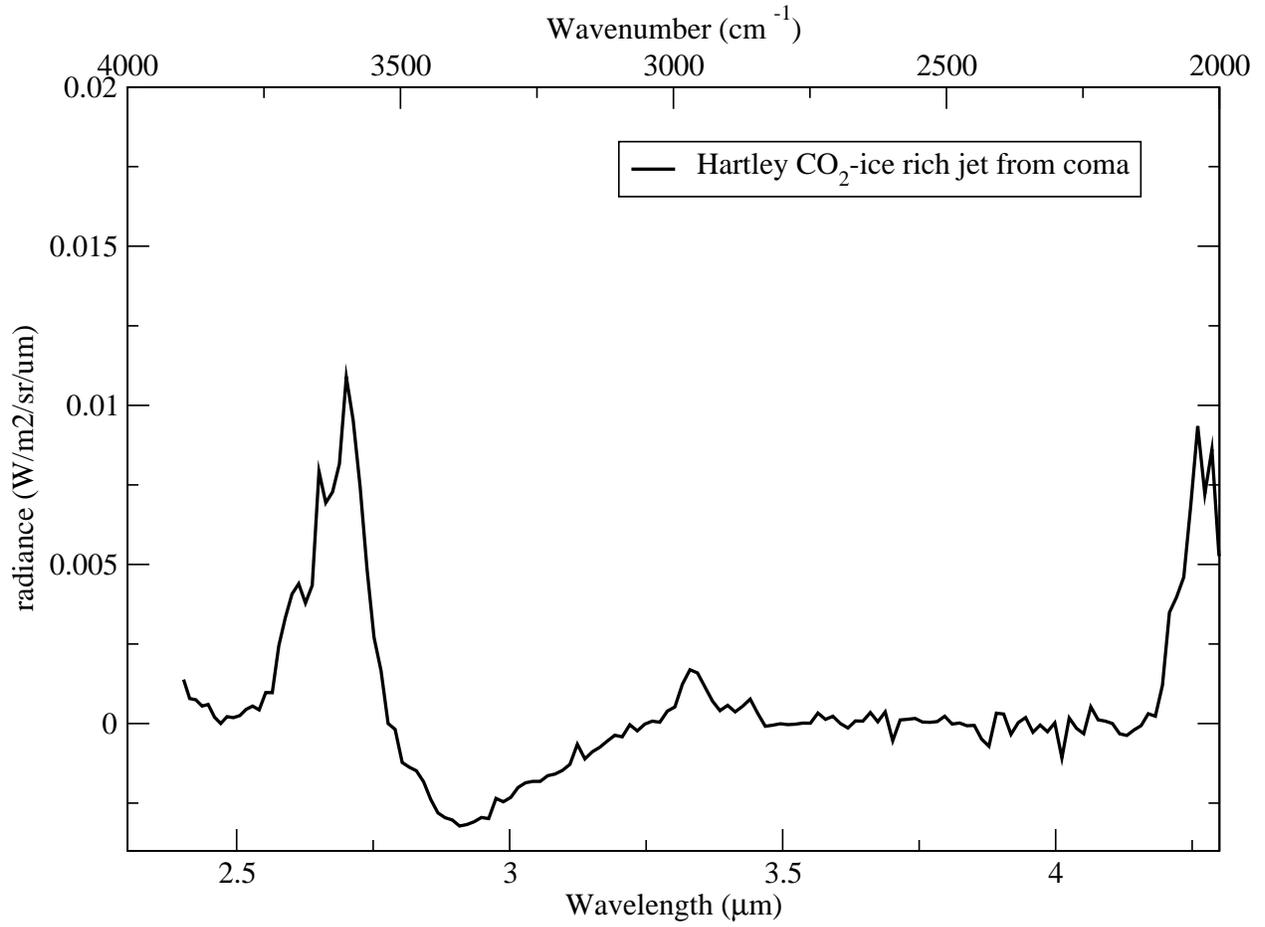}
\caption{IR spectrum of the coma of comet 103P/ Hartley 2.
The spectrum, reproduced from \citet{ahearn2011},
 was obtained from the CO$_{2}$-rich jet of the coma. \label{hartley-IR}}
\end{figure}

Organics were also detected in comet Hartley 2 \citep{ahearn2011, wooden2011}. In Figure \ref{hartley-IR} we have
 plotted, with permission from the authors, the spectrum obtained with the EPOXI space mission of the coma of
 comet 103P/Hartley 2 \citep{ahearn2011}. This is an ice-rich spectrum obtained from a CO$_{2}$ jet region of
 the coma \citep[see Fig. 5 of][]{ahearn2011}. From this spectrum we can see 3 major peaks in the 3.4 $\mu$m region
 at 3.336 $\mu$m (2997 cm$^{-1}$), 3.40 $\mu$m (2941 cm$^{-1}$) and 3.44 $\mu$m (2907 cm$^{-1}$). 
The relative depths and shapes of the
 peaks remain uncertain at this point, because the authors of this spectrum are still optimizing the calibration
 of the instrument, which will affect the continuum fit and therefore the absolute and relative flux level and
 shape of all of the features across all wavelengths (L. Feaga, personal communication). Therefore, our
 discussion will be limited to the peak positions only and no CH$_{2}$/CH$_{3}$ ratios will be calculated.
\citet{wooden2011} observed the coma of the comet and also obtained a spectrum, which is currently being processed
 for molecular methanol deconvolution and subtraction. The general profile of the 3.4 $\mu$m band with peak
 positions was communicated to us (Wooden, personal communication) but the CH$_{2}$/CH$_{3}$ ratio is not available
 yet because the optical depths are uncertain until they can properly subtract off the methanol molecular
 bands from the entire 3.4 $\mu$m feature. The spectrum they obtained (not shown) contains a broad band centered
 at 2938 cm$^{-1}$ and peaks at 2875, 2865, 2847 and 2829 cm$^{-1}$. 
The profile of the 3.4 $\mu$m band of both spectra of comet Hartley 2 looks very different from the 3.4 $\mu$m
 band in
 the DISM, 24 Themis, CCs and our samples, particularly the Wild 2 ones. From the spectrum shown in Figure
\ref{hartley-IR},
 we can see 3 main peaks at positions that are not found in any of our samples, the DISM and CCs. Only two of
 those peaks, at 2941 and 2997 cm$^{-1}$ (Table \ref{table3}) are also found in another object, 
asteroid 24 Themis. The third
 peak at 2906 cm$^{-1}$ is not found in any of the other objects and astrophysical environments we are comparing to. 
 Comet Hartley 2 seems to be dominated by compounds that absorb at 2940 cm$^{-1}$ and 2990 cm$^{-1}$ and whose
 nature is
 at present unknown. Similar compounds seem to be present in asteroid 24 Themis (Table \ref{table3}) but
 these compounds
 are absent in meteoritic materials and the DISM (Table \ref{table3}). As stated above, this comparison is purely
 qualitative given that the Hartley 2 spectra still need further processing and it may change once the
 deconvolution of the spectra is fully completed. Nevertheless, if the spectral features between the
 cometary spectra and our samples are indeed dissimilar, these differences suggest that the carrier of
 the 3.4 $\mu$m band in Hartley 2 is very different from the carrier of the 3.4 $\mu$m band in Wild 2, CCs,
 and DISM.
 Of particular interest is to note that two comets have such a different 3.4 $\mu$m band. While comet Wild 2 is
 dominated by CH$_{2}$ groups, comet Hartley 2 has only small CH$_{2}$ peaks and is dominated by
 an unknown compound.
 The non resemblance of the organics in these two comets suggest that there is either more than one way to
 manufacture organics in cometary regions or that the organics have been subsequently modified in their
 respective parent bodies. In any case more observations of comets are required to better constraint the 
nature and origin of the cometary organics responsible of the 3.4 $\mu$m band.

\subsubsection{The 3.4 $\mu$m feature in the Interstellar Medium}

\begin{figure}
\plotone{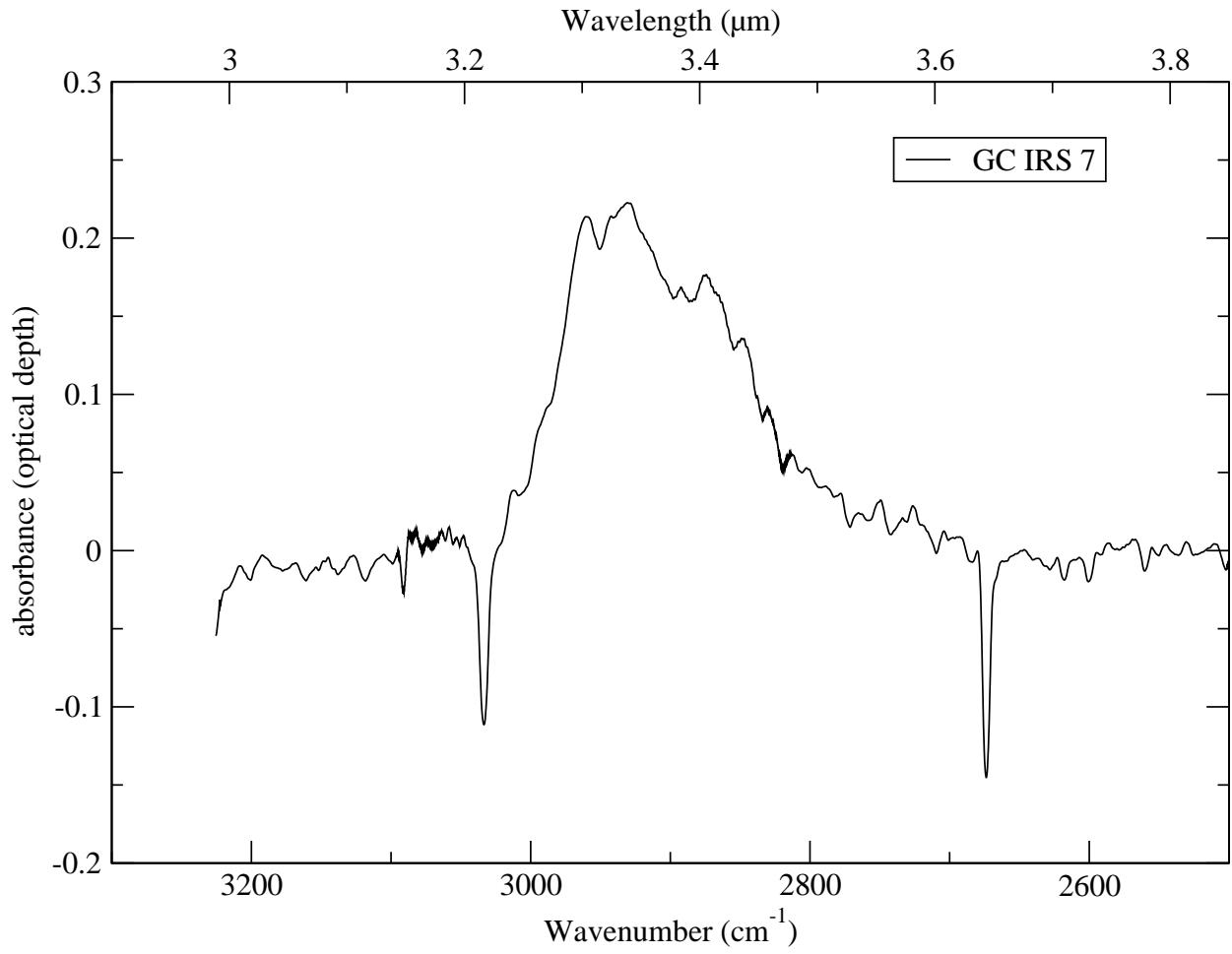}
\caption{IR spectrum of the galactic center source IRS 7,
obtained from
 the Infrared Space Observatory Data Center. \label{ISM-absorbance}}
\end{figure}

The spectrum of the source IRS 7 of the DISM (Figure \ref{ISM-absorbance}) was obtained from
 the Infrared Space Observatory Data Center and
 was reproduced  from \citet{matrajt2004}. This spectrum has several peaks in
 the 3000 cm$^{-1}$ region: 3012 cm$^{-1}$; 2959 cm$^{-1}$ (CH$_{3}$ asymmetric stretching); 
2928 cm$^{-1}$ (CH$_{2}$ asymmetric
 stretching); 2890 cm$^{-1}$ (CH$_{3}$ symmetric stretching); 
2872 cm$^{-1}$ (CH$_{3}$ symmetric stretching); 2848 cm$^{-1}$
 (CH$_{2}$ symmetric stretching). The band depth CH$_{2}$/CH$_{3}$ ratio 
was calculated to be between 1.1-1.4
 \citep{sandford1991, pendleton1994, matrajt2005}. Comparisons of IR spectra
 obtained from other lines of sight, including some extragalactic ones, have shown that the
 3.4 $\mu$m band has similar profiles 
\citep{pendleton1995, dartois2004, godard2012}.
 As discussed previously, the general shape of this band is quite different
 from the 3.4 $\mu$m band we observe in our samples. Nevertheless, many of these peaks
 are also found in other objects (Table \ref{table3}). As stated above for the spectrum of 24 Themis,
 the presence of all these peaks indicates that aliphatic chains are present as part of the
 compounds that make up for this band. The dissimilarity in the positions and relative
 strengths of the various peaks are probably due to differences in the relative abundances
 of -CH$_{2}$ and -CH$_{3}$ groups in the organic molecules of
the various objects and astrophysical environments we are comparing.
 For example, in the interstellar band the peaks corresponding to CH$_{3}$ symmetric stretching
 (2890 and 2872 cm$^{-1}$) are quite distinct whereas in the Wild 2 and IDP samples these are
 blended together in one unique peak at $\sim$2850 cm$^{-1}$ (Table \ref{table3}).
 The strength of the peak at 2959 cm$^{-1}$
 relative to the 2928 cm$^{-1}$ peak in the interstellar band when compared to the same peaks in
 our samples suggests that the interstellar carrier is richer in CH$_{3}$ groups than Wild 2 and
 IDPs. The CH$_{2}$ peaks in our samples are stronger and narrower than in the interstellar band.
 All these observations indicate that the aliphatic carriers in Wild 2 and IDPs are dominated
 by CH$_{2}$ groups (so they are mainly linear chains) whereas the carriers in the DISM are dominated
 by CH$_{3}$ groups (so they are ramified chains). 

\subsubsection{The 1700 cm$^{-1}$ (C=O) band in the DISM, IDPs and Wild 2 samples}

\begin{figure}
\plotone{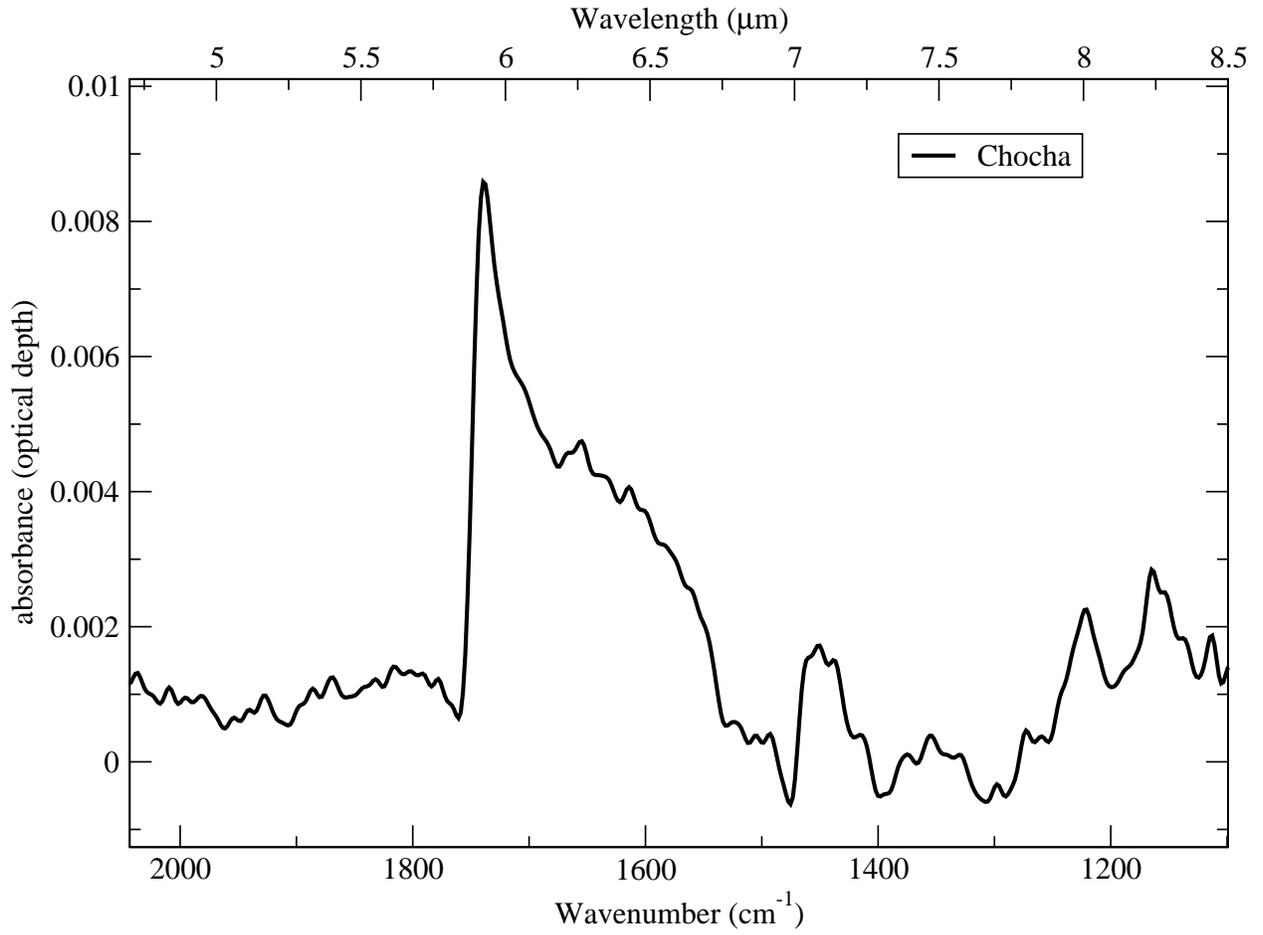}
\caption{IR spectrum of particle Chocha, zoomed in the
region around 1700 cm$^{-1}$. \label{Chocha-carboxyl}}
\end{figure}

\begin{figure}
\plotone{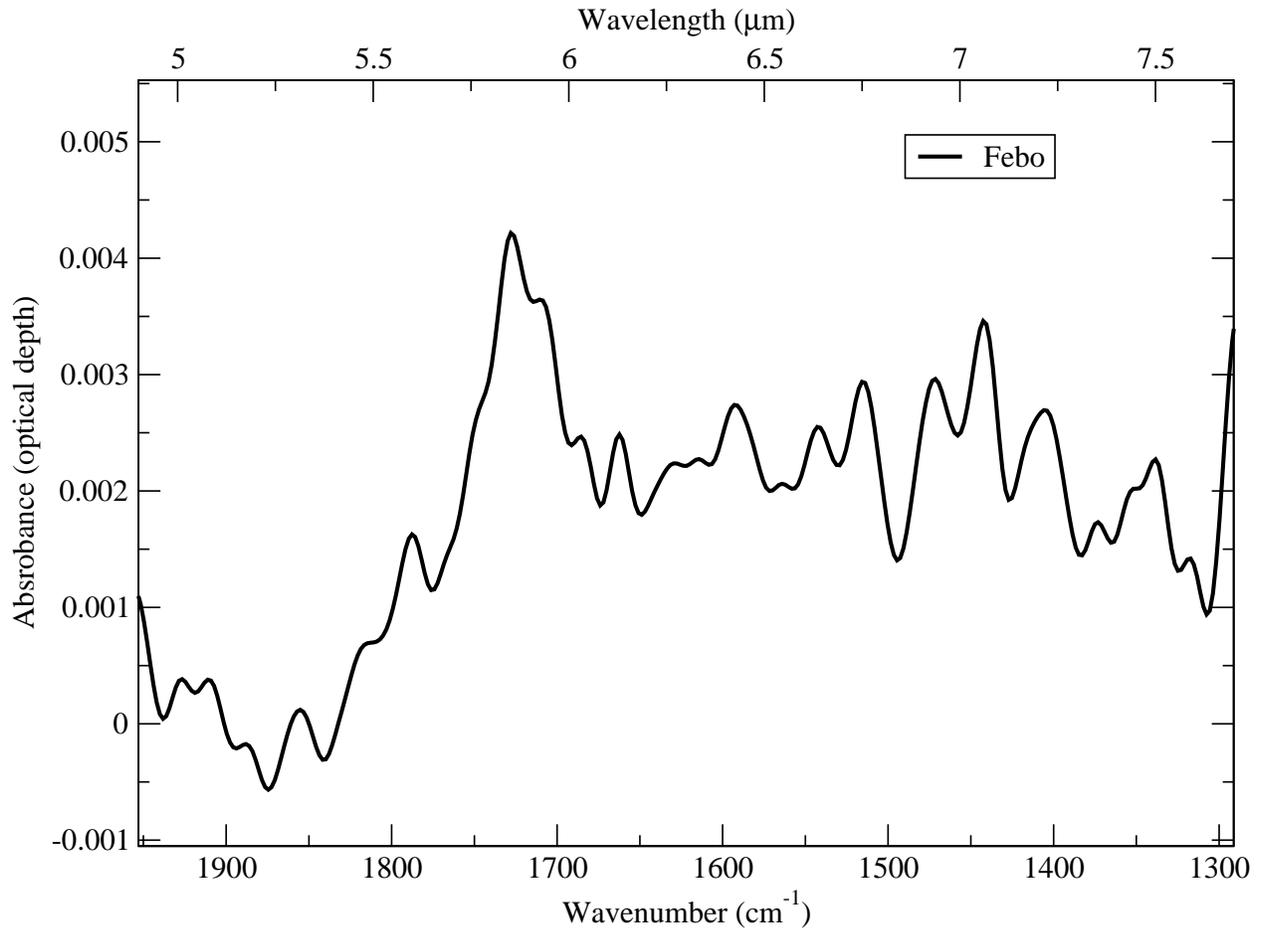}
\caption{IR spectrum of particle Febo, zoomed in the
region around 1700 cm$^{-1}$. \label{Febo-carboxyl}}
\end{figure}

\begin{figure}
\plotone{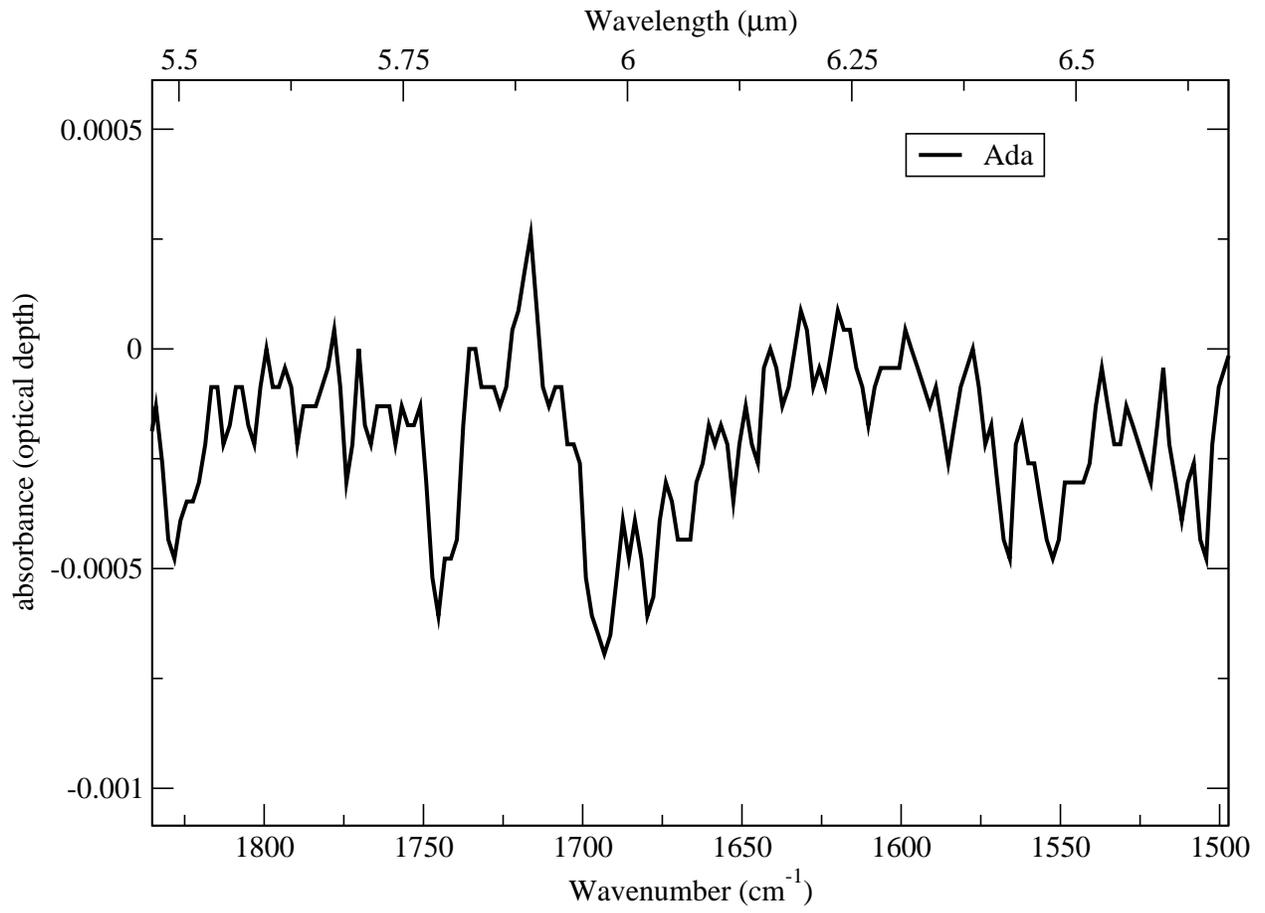}
\caption{IR spectrum of particle Ada, zoomed in the
region around 1700 cm$^{-1}$. \label{Ada-carboxyl}}
\end{figure}

The DISM of our galaxy is mainly composed of C and H with very little O  given
that no strong bands
 arising from other strong infrared active groups (like the C=O at 1700 cm$^{-1}$) are seen in
 the mid-infrared spectra \citep{dartois2004}. Only the extragalactic (Seyfert 2) NGC
 1068 mid-infrared spectrum displays a carbonyl absorption at 1703 cm$^{-1}$ \citep{dartois2004}.
 Most of our samples show a carbonyl peak (Table \ref{table3}).  For example, particles Chocha and Febo
 have a prominent peak at 1738 and  1727 cm$^{-1}$, respectively (Figures \ref{Chocha-carboxyl}
and \ref{Febo-carboxyl}). Sample Ada has a small
 peak at 1715 cm$^{-1}$ (Figure \ref{Ada-carboxyl}). Particle GS does not have a C=O peak. A carbonyl peak has been
 detected in other IDPs in previous IR studies \citep{flynn2003, matrajt2005}.
 Furthermore, studies with other analytical techniques, for example X-ray microscopy, have shown
 that IDPs \citep{flynn2003} and Wild 2 samples \citep{matrajt2008,
degregorio2010, degregorio2011}
 have a C=O peak associated with the organic materials. The presence of a carbonyl group on most
 IDPs and Wild 2 and its absence in the DISM of the Galaxy is a further indication that the
 organic materials in IDPs and Wild 2 are quite different from the interstellar organics,
 strongly suggesting that they did not originate in the DISM (see next section).

\subsection{Comparison of the samples spectra to modeled spectra}

\citet{jones2012I} modified the eRCN and DG models to better constraint and determine the
 infrared spectral properties of amorphous
hydrocarbon grains (also known as hydrogenated amorphous carbons HAC) found in the ISM, as a function of
their content in hydrogen atoms (noted X$_{H}$). In this model, it was found that when the molecules
suffer from structural annealing, they lose H atoms and transform from an aliphatic-rich structure to
an aromatic-rich structure. This annealing would be UV photon-driven from the radiation field in the
ISM environment. However, this annealing would highly depend on the size of the particles being
exposed to the radiation and thermal effects \citep{jones2012III}. In particular, for carbonaceous
grains $>$ 3 nm in radius the radiation effects are negligible \citep{jones2012III} and so their
IR spectra remain invariant. It is very interesting to compare our spectra to the spectra predicted by this model,
because although our grains are typically $>$ 50 nm, the particles suffered from heating effects either
during atmospheric entry or during deceleration into aerogel. In addition, the model predicts how the spectra should look as a function
of the CH$_{2}$/CH$_{3}$ ratios calculated for the carbonaceous materials, because these ratios determine the atomic content of hydrogen (X$_{H}$)
in the molecules.

Using the CH$_{2}$/CH$_{3}$ ratios calculated for each of our 4 samples (Table \ref{table2})
and using the graph of figure 3 from \citet{jones2012I}, we determined the X$_{H}$ for each of our samples.
Then, with the corresponding X$_{H}$ we searched for the corresponding spectrum predicted by the model
(figures 8-10 from \citet{jones2012I}) and compared them. Particles Chocha and Ada have a CH$_{2}$/CH$_{3}$
equivalent to a X$_{H}$ of $\sim$ 0.53 (meaning the carbonaceous material has a hydrogen
atom fraction of 53 $\%$). The predicted spectrum obtained with such hydrogen fraction
looks quite similar to our spectra. First, the intensity of the CH$_{3}$ peak is very small relative to the intensity
of the CH$_{2}$ peak. Second, the spectrum does not seem to have peaks in the 6-6.67 $\mu$m range (1600-1500 cm$^{-1}$),
typically of CH aromatic moieties. Our spectra do not have peaks in that region either. Third, the predicted spectrum
does not have peaks in the 3.28 $\mu$m region (3050 cm$^{-1}$) typically of aromatics. Instead the spectrum 
has a nice little peak at 3.25 $\mu$m and a nice sharp peak at 3.32 $\mu$m (3078 and 3010 cm$^{-1}$, respectively)
corresponding to alkene (olefin C=C) groups. Our spectra do not have any type of aromatic or olefin
peaks in this region, which makes the biggest difference between the predicted spectrum and our observed spectra.
But the overall profile of the 3.4 $\mu$m band matches up quite well between the predicted spectrum and our
observed spectra. We think that the small discrepancies are related to differences in the composition
of the carbonaceous materials of our samples. For example, we have determined previously
that these carbonaceous materials all contain bonded O and N heteroatoms \citep{matrajt2008,matrajt2012}.
The eRCN and DG models of \citet{jones2012I} do not take into account the presence of heteroatoms.
In addition, the model predicts that aromatics tend to increase as aliphatics decrease under thermal effects
but only for particles $<$ 10 nm, and as we previously stated, all of our materials are $>$ 50 nm.
Finally, it is worth noting that this model presents discrepancies even with the ISM
spectra, where the CH$_{2}$/CH$_{3}$ ratios range between 2-3 (equivalent to 53-55 $\%$ of H atoms),
corresponding to a predicted spectrum with much smaller CH$_{3}$ than the one observed \citep{jones2012I}.
To obtain a perfect match, one would need to increase the intensity of the
CH$_{3}$ band in order to obtain a good match to the observational
data. We also noticed that all our spectra show an overall decrease in the intensity of the CH$_{3}$ peak
relative to the ISM spectra, and we think that this is a fundamental difference between the carbonaceous
material of the ISM and our samples.

Regarding the other two samples, Febo and GS, the comparison is actually quite different.
The CH$_{2}$/CH$_{3}$ ratios of these two particles correspond, respectively, to
a X$_{H}$ of 0.55 and 0.58. The predicted spectra obtained with these
hydrogen atomic fractions do not match our spectra. They
seem to have much more CH$_{2}$ than our spectra. In addition, the relative intensities
of the CH$_{3}$ and the CH$_{2}$ peaks are quite different than the ones we observe in our spectra.
Finally, the predicted spectrum shows some olefinic peaks that are absent in our spectra.
As stated above, we think that some of the discrepancies encountered while doing this comparison
are related to the presence of heteroatoms in our samples and the size of the carbonaceous grains.

This model predicts that 
sp$^{3}$ C atoms (aliphatics) are transformed into sp$^{2}$ C atoms (aromatics) as annealing
proceeds. This could in principle tells us something about the effects of heating during atmospheric
 entry and impact into aerogel. However, as discussed above, what we observe is the exact opposite: we do not observe 
 aromatics or olefinic material. Instead,  the carbonaceous materials of our samples are dominated by aliphatics.
Therefore, we think that, as stated by the model, the annealing effects are valid for a certain size,
but because the carbonaceous materials of our samples are $> 50$nm, we do not observe those effects.

\subsection{Is the Origin of the 3.4 $\mu$m band in IDPs and Wild 2  interstellar?}
\subsubsection{IDPs}

We compared the 3.4 $\mu$m band in the two IDPs we studied with the 3.4 $\mu$m
 band found in IDPs of previous studies \citep{flynn2003, matrajt2005}.
 We found that our IDPs have a similar profile to most
 of the IDPs we studied previously  in
 which there is clear dominance of CH$_{2}$ over CH$_{3}$,
 sometimes with CH$_{3}$
 almost nonexistent (for example Chocha in this study and samples
 L2011-F2, L2036-E19, W7116B-N and W7116B-U2 from \citet{matrajt2005}).
 However, this profile is quite different from the 3.4 $\mu$m band of the
 DISM, where there is not a marked dominance of CH$_{2}$ groups over CH$_{3}$
 groups as seen in IDPs. Instead, in the 3.4 $\mu$m band of the DISM  the
 CH$_{2}$ and CH$_{3}$ features are comparable, whereas in all the spectra of the
 IDPs we studied previously, the CH$_{2}$ tends to dominate 
\citep{flynn2003, matrajt2005}.
 In addition, the band depth ratios CH$_{2}$/CH$_{3}$ found
 for IDPs previously studied ranges from 1.88-3.69 with an average of
 2.47 \citep{matrajt2005} and from 1.0-5.6 with an average of 2.4
 \citep{flynn2003}. The band depth CH$_{2}$/CH$_{3}$ ratio 1.01 found for particle
 GS (Table \ref{table2}) is in good agreement with the ratios found previously and it
 is also within the range of values found for the DISM and for the IOM of
  CCs (Table \ref{table2}). However, the profile of the 3.4 $\mu$m band is quite different
 from the DISM and the IOM of  CCs. Therefore, even if the ratios are similar
 the profiles are not, which suggests that the organic compounds that make up
 the 3.4 $\mu$m band are not related to the DISM and the parent body of GS is not
 a carbonaceous chondrite parent body. The band depth ratio CH$_{2}$/CH$_{3}$ of sample
 Chocha is 4.6 (Table \ref{table2}), still in the range previously found for IDPs but
 about 4 times bigger than the band depth CH$_{2}$/CH$_{3}$ ratio found for the DISM
 (1.1-1.25, \citet{sandford1991}). Both the ratio and the profile of particle
 Chocha are considerably different from the DISM and the carbonaceous chondrites
 IOM, suggesting that the parent body of Chocha is not a CC. The organic material
 of Chocha has isotopic anomalies indicative of an origin in a cold environment
 which could be either the ISM or the edges of the protoplanetary disk 
\citep{matrajt2012}.
 Given that the 3.4 $\mu$m band seems quite different from the DISM as
 discussed above, we suggest that the origin of the carbonaceous materials in
 Chocha is not interstellar but in the edges of the protoplanetary disk.
The other IDPs we studied here and in the past also show a 3.4 $\mu$m band with
 a different profile, although for some of them the CH$_{2}$/CH$_{3}$ ratios are similar
 to the DISM. The average is, however, twice as big as the ratio found for the
 DISM and IOM of CCs (Table \ref{table2}). This indicates, as previously observed
 \citep{sandford2006, munozcaro2008}, that the aliphatic molecules
 in most IDPs are longer (or less branched) than those in the DISM and suggest
 that they are not similar to the organic materials responsible for the 3.4 $\mu$m
 feature in the DISM, which looks more similar to the insoluble organic matter
 found in Orgueil and Murchison (Table \ref{table2}). These conclusions are further
 supported by the presence of O (detected by XANES, \citet{flynn2003}), the
 presence of N (detected by XANES, \citet{keller2004} and by NanoSIMS, 
\citet{matrajt2012}) and the presence of the C=O (Table \ref{table3}, Figure
\ref{Chocha-absorbance})
 in many of these IDPs. We conclude that the origin of the 3.4 $\mu$m band in
 IDPs is not related to interstellar material. We also conclude that the
 organic materials making up this band are not related to the parent bodies
 of CCs. Based on all our observations discussed above we conclude that the
 origin of the 3.4 $\mu$m band in IDPs is most likely in the outer parts of the
 solar nebula. These results are in agreement with recent models of the dynamics
 and evolution of icy grains exposed to UV irradiation and the production of
 complex organics, which have shown that organic compounds are natural byproducts
 during the evolution of the protoplanetary disk \citep{ciesla2012}.

\subsubsection{Wild 2}
We compared the 3.4 $\mu$m band of our two Wild 2 samples with the 3.4 $\mu$m band of Wild 2 
samples studied in the past \citep{sandford2006, keller2006,
munozcaro2008}
  and with the DISM. The profile of the 3.4 $\mu$m band of both of our Wild 2 samples
 is quite similar, with little to no CH$_{3}$ features and a clear dominance of CH$_{2}$ features
 with strong, narrow peaks. This profile resembles the profiles of other Wild 2 samples
 studied in the past \citep{sandford2006, keller2006,
munozcaro2008}. 
 The band depth ratio CH$_{2}$/CH$_{3}$ of particle Febo is 1.96. This value is twice as big as
 the ratio in DISM and IOM of  CCs. The band is dominated by CH$_{2}$ groups. Similarly,
 the band depth CH$_{2}$/CH$_{3}$ ratio of particle Ada (4.3) is much bigger than the ratio found
 in the DISM and IOM of CCs. The profile of the band is also dominated by CH$_{2}$ groups.
 This indicates that the aliphatic organic materials responsible of this band have longer
 or less branched chains than those in IOM of  CCs and the DISM. In addition, we found
 the presence of the carbonyl group (C=O, Table \ref{table3}, Figures \ref{Febo-carboxyl}
 and \ref{Ada-carboxyl}) and previous studies
 have established that the organic materials in Wild 2 contain O and N 
\citep{sandford2006, matrajt2008, degregorio2010, matrajt2013}.
 \citet{munozcaro2008} 
  compared the 3.4 $\mu$m feature of several Wild 2 samples to ice residues obtained by
 UV-photoprocessing of interstellar ice analogs and showed that the 3.4 $\mu$m band in Wild 2
 samples is spectroscopically different from the analog organic residues. This indicates
 that the aliphatic organics that produced the 3.4 $\mu$m band are not a direct product of
 interstellar ice photoprocessing. Isotopic compositions obtained from the two Wild 2
 particles analyzed in our study \citep{matrajt2008} and from other Wild 2 particles
 \citep{degregorio2010, degregorio2011} showed $^{15}$N anomalies indicative of an origin in a cold
 environment. All these observations  and the discussions presented above indicate that
 the DISM is not a good candidate for the synthesis of the aliphatics that make up the 3.4
 $\mu$m band in Wild 2 samples. Therefore we conclude that, similar to
IDPs, the origin of the 3.4 $\mu$m band in Wild 2 particles is most likely
 in the outer parts of the solar nebula. Our conclusions are in
 agreement with previous statements that the Wild 2 organic materials
 are not the direct result of diffuse ISM processes but rather result
 from protosolar nebular processes \citep{sandford2006}.

\section{Conclusions}

This study showed that both IDPs and Wild 2 samples have a 3.4 $\mu$m
 band dominated by CH$_{2}$ groups, indicating that the organic compounds
 are predominantly long aliphatic chains with little ramifications.
 Previous studies showed that all these samples have O and N bonded
 to their organic materials, suggesting that these ramifications are O
 and N-rich groups. Most of our samples have the C=O carbonyl features
 between 1000-2000 cm$^{-1}$. The comparison of all our samples to the 3.4 $\mu$m
 band of the DISM and IOM of carbonaceous chondrites showed that the
 profile of the bands in Wild 2 and IDPs is quite different from the
 DISM and CCs because the band is mainly dominated by CH$_{2}$ groups.
 In some of our samples the CH$_{3}$ groups are absent while in the DISM
 and IOM of CCs the CH$_{3}$ groups are of comparable abundance with the CH$_{2}$
 groups. This indicates that our samples are richer in CH$_{2}$ than the
 DISM. All these observations suggest that the organic component in
 Wild 2 and IDPs does not have an interstellar origin. We conclude
 that the organic compounds in both IDPs and Wild 2 were formed at
 the edges of the protoplanetary disk through nebular processes.
 It is possible, however, that a pristine interstellar component
 lies underneath a more complex organic material that formed during
 solar nebula processing or later in the parent body of Wild 2 and IDPs.
 We favor, however, the scenario in which the organic materials in both
 Wild 2 samples and IDPs originated at the edges of the protoplanetary disk.

\section{Acknowledgments}

We are grateful to H. Campins and colleagues and to A. Rivkin and collaborators
 for having shared with us their 24 Themis spectra. We are also grateful to
M. A$\textquoteright$Hearn and collaborators and to D. Wooden for having
shared their spectra of comet Hartley 2. We also thank H. Bechtel for his assistance.
We would like to thank Anthony Jones, whose review and comments very much improved
this article. This work is dedicated to the memory of Carl Sagan (1934-1996) who
greatly inspired G.M throughout her career.
G. M was funded by NASA grants  NNX10AI89GS01 and NNG06GG00GS05. 

{\it Facilities:} \facility{University of Washington}, \facility{Brookhaven National Lab.}, 
\facility{Lawrence Berkeley National Lab}.



\clearpage

\clearpage

\clearpage

\clearpage

\clearpage

\clearpage

\clearpage

\clearpage

\clearpage

\clearpage

\clearpage

\clearpage

\clearpage

\clearpage

\clearpage

\clearpage

\clearpage

\clearpage

\clearpage

\clearpage

\clearpage







\clearpage

\begin{deluxetable}{cccc}
\tabletypesize{\scriptsize}

\tablecaption{Peak assignments. The assignments
 are based on what was previously reported in the litterature 
\citep[and references therein]{matrajt2004, matrajt2005, munozcaro2006}. \label{table1}}
\tablehead{
\colhead{Peak position (in cm$^{-1}$}) & \colhead{Vibration mode} & \colhead{Interpretation} 
& \colhead{sample that has it}
}
\startdata
3255, 3270 & OH & water & GS, Chocha   \\
2990 & C=C-H & - & acrylic   \\
2951,2954,2950,2949,2958 & CH$_{3}$ asymmetric stretching & aliphatic hydrocarbons & GS, Chocha, Febo, Ada, Acryic, DISM \\
2920, 2918, 2929, 2925, 2922 & CH$_{2}$ asymmetric stretching & aliphatic hydrocarbons & GS, Chocha, Febo, Ada, DISM \\
2896, 2870, 2860, 2874 & CH$_{3}$ symmetric stretching & aliphatic hydrocarbons & GS, Febo, DISM \\
2845, 2847, 2855 & CH$_{2}$ symmetric stretching & aliphatic hydrocarbons & GS, Chocha, Febo, Ada, DISM \\
2160 & C=C stretching & - & Ada \\
1740 & C=O carbonyl & esters & Chocha \\
1730, 1717, 1700, 1714, 1727 & C=O carbonyl & ketone, carboxylic acid & Febo, Ada, acrylic \\
1685 & H-O-H & water & Febo \\
1654, 1650 & C=C stretching & aromatics & Chocha, Febo, Ada, acrylic \\
1545-1455 & CO$^{3-}$ & carbonates & GS \\
1480 & CH$_{3}$ asymmetric bending & aliphatic hydrocarbons & Chocha, acrylic \\
1447-1448 & CH$_{2}$ asymmetric bending & aliphatic hydrocarbons & Chocha, acrylic \\
1418, 1435 & C=C stretching & aromatics & Chocha, Febo, Ada, acrylic \\
1350, 1386 & CH$_{3}$ symmetric bending & aliphatic hydrocarbons & acrylic \\
1240, 1270 & C-O-C & esters & acrylic \\
1220 & CH$_{2}$ symmetric bending & aliphatic hydrocarbons & Chocha \\
1147, 1190 & unknown & - & - \\
1160 & CH$_{2}$ twisting & aliphatic hydrocarbons & Chocha \\
1065 & C-OH & secondary cyclic alcohols & acrylic \\
987, 970, 910 & CH=CH bending & - & acrylic \\
1070, 1060, 952 & Si-O & pyroxene & GS, Febo \\
1216, 1136, 1106, 1010, 930 & Si-O & silicates & Febo, Ada \\
880 & Si-O & olivine & Febo \\
\enddata

\end{deluxetable}

\clearpage

\begin{deluxetable}{ccc}
\tabletypesize{\scriptsize}

\tablecaption{CH$_{2}$/CH$_{3}$ band depth ratios. \label{table2}}
\tablehead{
\colhead{Object/sample} & \colhead{CH$_{2}$/CH$_{3}$} & \colhead{Reference}
}
\startdata
DISM (GC IRS7) & 0.96-1.25 & \citep{sandford1991}   \\
DISM (GC IRS7) & 0.92-1.2 average 1.06 & \citep{pendleton1994} \\
Extragalactic ISM (Seyfert 2) & 2.0 & \citep{dartois2004} \\
Tagish Lake & 4.36 & \citep{matrajt2004} \\
IDPs & 1.0-5.6 average 2.4 & \citep{flynn2003} \\
IDPs & 1.88-3.69 average 2.47 & \citep{matrajt2005} \\
Wild 2 samples & 1.7-2.8 average 2.15 aerogel 2.15 & \citep{munozcaro2008} \\
Wild 2 samples & 2.5 & \citep{keller2006, sandford2006} \\
IOM Murchison & 1.5 & \citep{ehrenfreund1991} \\
IOM Murchsion & 1.09 & \citep{flynn2003} \\
IOM Orgueil & 1-1.51 & \citep{ehrenfreund1991} \\
ultracarbonaceous IDP Chocha & 4.6 & this study \\
ultracarbonaceous IDP GS & 1.01 & this study \\
Wild 2 Febo & 1.96 & this study \\
Wild 2 Ada & 4.3 & this study \\
Comet 103P/Hartley 2 (coma) & no CH$_{3}$ or CH$_{2}$ bands so no ratio calculated & \citep{ahearn2011, wooden2011} \\
Asteroid 24 Themis & not calculated & \citep{campins2010, rivkin2010} \\
\enddata

\end{deluxetable}

\clearpage

\begin{deluxetable}{cccccccccccc}
\tabletypesize{\scriptsize}
\rotate

\tablecaption{Comparison of the 3.4 $\mu$m band peaks and the C=O peak
between different objects \label{table3}}
\tablewidth{0pt}
\tablehead{
\colhead{Peak (cm$^{-1}$)} & \colhead{GS} & \colhead{Chocha} & \colhead{Febo} & \colhead{Ada} &
\colhead{DISM} & \colhead{Seyfert 2} & \colhead{IOM Murchison} & \colhead{IOM Orgueil} &
\colhead{Tagish Lake} & \colhead{24 Themis} & \colhead{Hartley 2} \\
}
\startdata
3250-2670 (OH)       & yes & yes & no  & no  & no  & -   & yes & yes & yes & yes & -   \\
2950-2955 (CH$_{3}$) & yes & yes & yes & yes & yes & yes & yes & yes & yes & yes & yes \\
2918-2925 (CH$_{2}$) & yes & yes & yes & yes & yes & yes & yes & yes & yes & yes & no \\
2865-2896 (CH$_{3}$) & yes & no  & no  & no  & blended& yes & yes& yes & yes & yes & yes \\
2845-2855 (CH$_{2}$) & yes & yes & yes & yes & blended& no  & no & no  & yes & yes & yes \\
1700 (C=O)           & no &  yes & yes & yes & no     & yes & yes & -  & no  & -   & -  \\
\enddata


\end{deluxetable}

\clearpage
\bibliographystyle{apj}
\bibliography{FTIR2}
\end{document}